\documentclass[11pt,a4paper]{article}

\def\bSig\mathbf{\Sigma}

\usepackage{amsmath,tikz}
\usetikzlibrary{matrix}
\usepackage{amsfonts}
\usepackage{amssymb}
\usepackage{bbm}
\usepackage{xcolor}
\usepackage{hyperref}
\usepackage{natbib}
\usepackage{subcaption}
\usepackage{pifont}
\usepackage{setspace}
\usepackage{ulem}

\usepackage[top=1.0in, bottom=1.0in, left=1in, right=1in]{geometry}
\usepackage[labelformat=parens,labelsep=quad,skip=4pt,
font=small,format=plain, labelfont=it, textfont=it]{caption}

\newtheorem{Rem}{Remark}[section]

\newtheorem{The}{Theorem}[section]
\newtheorem{Lem}{Lemma}[section]

\newenvironment{proof} {\noindent {\textbf{Proof}}} { \hfill $\Box$ \\ }
\newtheorem{Prop}{Proposition}[section]

\newcommand{\R}{\mathbb{R}}

\newcommand{\E}{{\mathbb E}}

\newcommand\norm[1]{\left\lVert#1\right\rVert}

\usepackage{comment}
\usepackage{rotating}

\title{\vspace{-50pt}Bootstrap inference for linear regression\vspace{-15pt} between variables that are never jointly observed: application in in vivo experiments}

\author{Polina Arsenteva$^{1,2,*}$, 
Mohamed Amine Benadjaoud$^{3}$, and Herv\'{e} Cardot$^{1}$ \\
\footnotesize{ \vspace{-15pt} $^{1}$Institut de Math\'{e}matiques de Bourgogne, UMR CNRS 5584, Universit\'{e} Bourgogne Europe, Dijon, France} \\
\footnotesize{ \vspace{-15pt} $^{2}$IRSN PSE-SANTE/SERAMED/LRMed, Fontenay aux roses, France} \\
\footnotesize{$^{3}$IRSN PSE-SANTE/SERAMED/LRAcc, Fontenay aux roses, France}}

\date{}

\begin{document}

\maketitle

\begin{abstract}
	In modern experimental science, there is a common problem of estimating the coefficients of a linear regression in a context where the variables of interest cannot be observed simultaneously. When there is a categorical variable that is observed on all statistical units, we consider two estimators of linear regression that take this additional information into account: an estimator based on moments and an estimator based on optimal transport theory. These estimators are shown to be consistent and asymptotically Gaussian under weak hypotheses. The asymptotic variance has no explicit expression, except in some special cases, for which reason a stratified bootstrap approach is developed to construct confidence intervals for the estimated parameters, whose consistency is also shown. A simulation study evaluating and comparing the finite sample performance of these estimators demonstrates the advantages of the bootstrap approach in several realistic scenarios. An application to in vivo experiments, conducted in the context of studying radio-induced adverse effects in mice, revealed important relationships between the biomarkers of interest that could not be identified with the considered naive approach.
\end{abstract}

\section{Introduction}\label{sec:intro}
In vivo experiments are often used to study the effects of a treatment on a living organism. In the context of a complex organism response, scientists may be interested in studying multiple variables that describe the effect at different scales. In particular, such variables of interest often include a macroscopic biomarker that is only available through in vivo data, and a microscopic biomarker that can also be observed at the cellular level (i.e. in vitro). The interest in this case is to use the latter to predict the former. For example, in the context of studying the adverse effects of radiotherapy on healthy tissue, the potential outcomes of interest are the severity of macroscopic lesions and predictors such as gene expression. In preclinical research, these quantities of interest are often observed in different animals from independent cohorts. Since the goal is to establish relationships between these variables, the problem of statistical data fusion arises.

\begin{figure}
	\centering
	\includegraphics[clip, trim=12cm 8cm 12cm 13cm,width=0.9\textwidth]{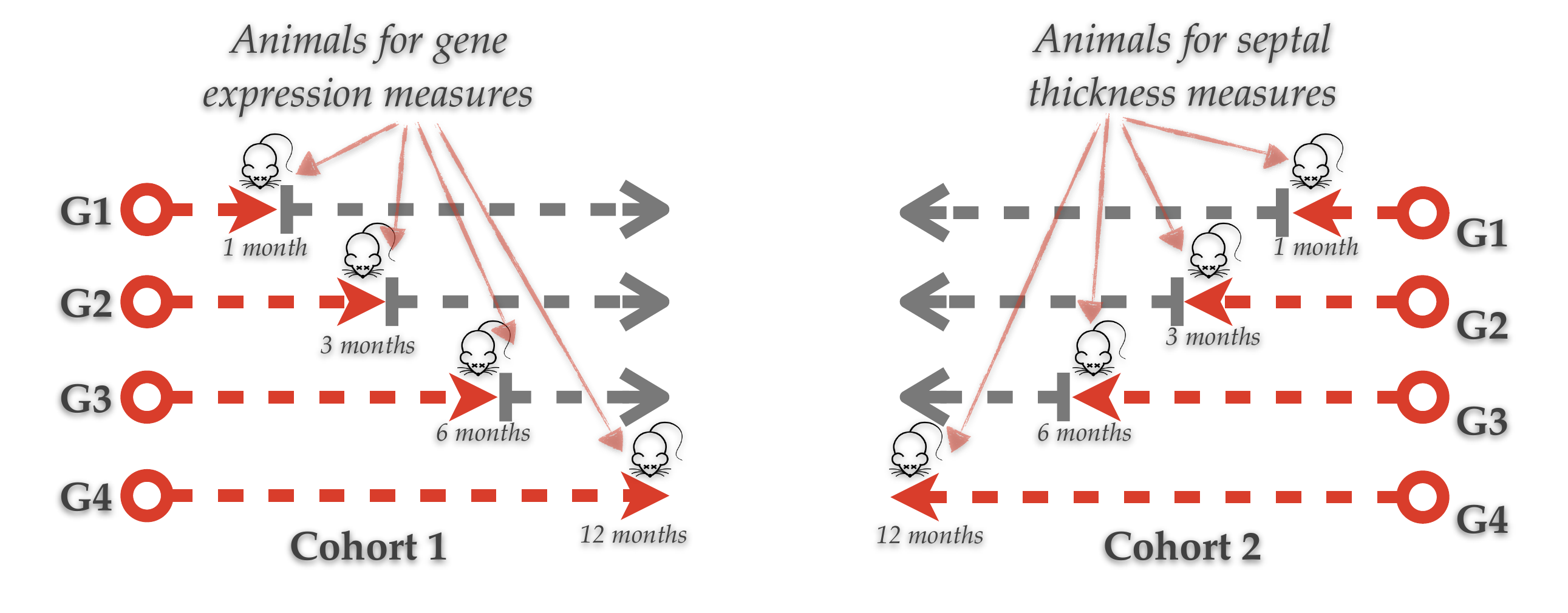}
	\caption{\label{fig:vivo_exp_ex}Schematic representation of the design of an in vivo experiment studying the effect of irradiated volume.}
\end{figure}

An illustrative example of an in vivo experiment where the variables of interest are not observed simultaneously is shown in Figure \ref{fig:vivo_exp_ex}. In this experiment, presented in \cite{bertho_preclinical_2020}, mice are irradiated in the lungs with different volumes to study the role of irradiated volume in the occurrence of radiation-induced adverse effects. The latter are assessed by measuring septal thickening, a histological marker of lung injury. The other variable measured to predict the adverse effect variable is the expression of several pro-inflammatory genes. As shown in Figure \ref{fig:vivo_exp_ex}, there are two independent cohorts in the study, one used to measure gene expression and the other to measure septal thickness.

\begin{figure}
	\centering
	\includegraphics[clip, trim=0cm 0cm 0cm 1cm, width=0.8\textwidth]{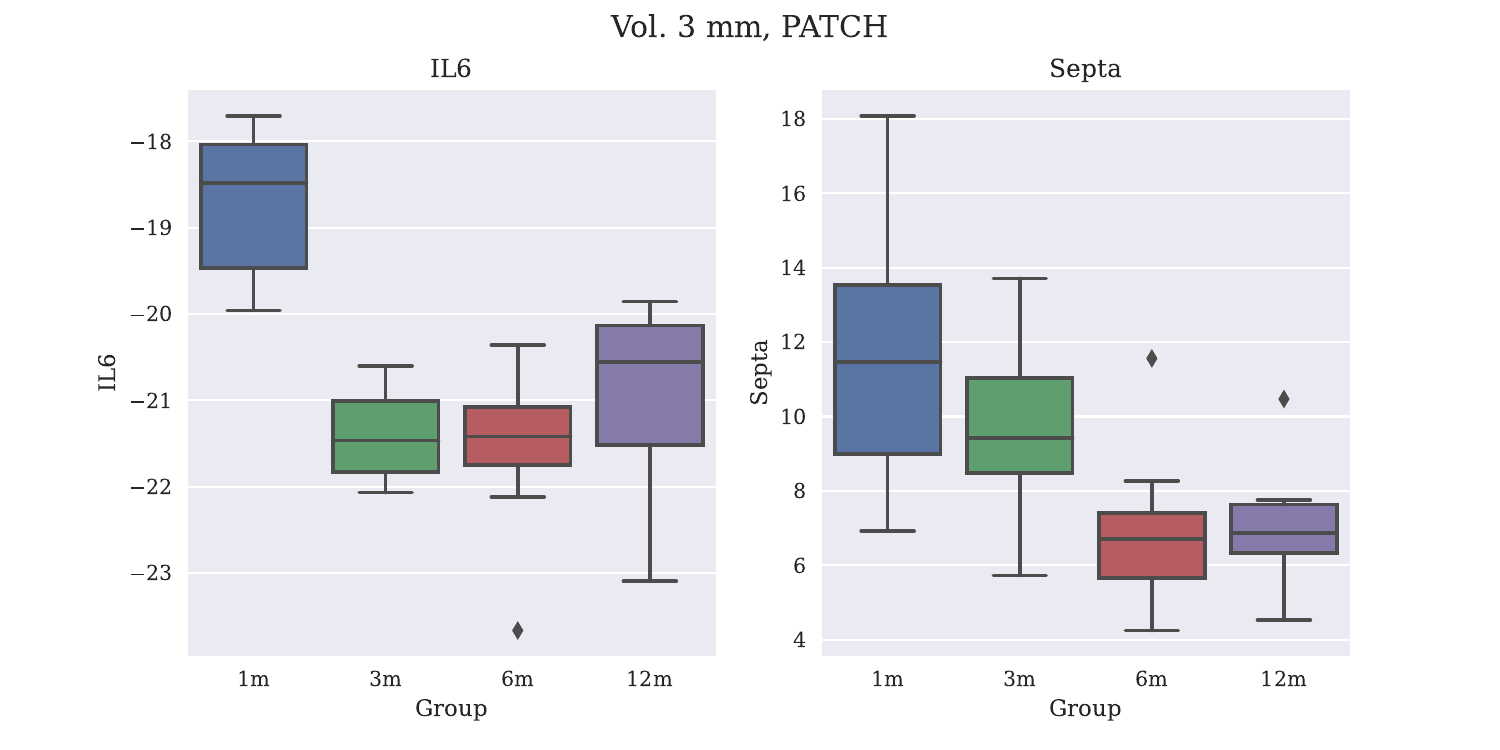}
	\caption{\label{fig:vivo_data_ex}Distribution of the data, collected from the irradiated patch under SBRT with 3 mm beam size: the expression of the gene IL6 on the left, and septal thickness on the right. The measurements were made 1, 3, 6 and 12 months after irradiation.}
\end{figure}

Comparing the distributions of measurements from the two cohorts, as shown in Figure \ref{fig:vivo_data_ex}, may suggest a correlation or even a linear relationship between the variables. To establish whether such a relationship exists, it is necessary to link two variables that are not observed on the same statistical units, which is equivalent to solving a data fusion or a file matching problem according to the terminology employed in  \cite{Little_Rubin_2002}. In this example, there are four groups indicating the time points (1, 3, 6 and 12 months after irradiation) when the measurements were taken on the corresponding animals. Thus, the categorical variable indicating the time point, which is known for each observation, can be used as an additional variable to link the predictor and the predicted variables.

The task of linking variables that are not observed together cannot be approached as a typical missing value problem, since most methods of inference on incomplete data require sufficient overlap, which is completely absent in the case we are dealing with. As a result, all approaches that use frameworks such as multiple imputation in the context of data fusion are inappropriate for our application. For example, \cite{carrig_nonparametric_2015} use multiple imputation to integrate different datasets, which allows for the absence of overlap, but requires a calibration dataset in which all variables of interest must be jointly observed. 

Other approaches to data fusion available in the literature include factor analysis \citep{cudeck_estimate_2000}, statistical matching \citep{mitsuhiro_kernel_2020}, Bayesian network inference \citep{triantafillou_learning_2010, tsamardinos_towards_2012} and Gaussian Markov combinations \citep{massa_algebraic_2017}. These methods are designed to link variables that are not simultaneously observed through covariates that exist for both variables of interest. This corresponds to the characteristics of in vivo data described above. However, the covariates in these approaches are continuous random variables,
often assumed to be Gaussian, as is the case in \cite{cudeck_estimate_2000} and \cite{massa_algebraic_2017}. The grouping variable available from in vivo experiments cannot be represented in continuous form, since categories such as control and sham make it impossible to assume continuity and normality. The Bayesian network approaches introduced by \cite{triantafillou_learning_2010} and \cite{tsamardinos_towards_2012}, which aim to infer binary causal relationships between variables, are more suitable for large datasets with a high number of covariates. Current research in statistical matching addresses aspects such as not-at-random missingness \citep{mitsuhiro_bayesian_2021} and high dimensionality \citep{mitsuhiro_kernel_2020}. This approach is based on the idea of comparing distances between covariates from the datasets of interest, which cannot be done by taking the group variable as a covariate. It can be noted that the goal of the aforementioned examples in statistical matching is to group individuals prior to imputation, which is not necessary in our case since the groups are already known. Finally, ecological regression-based approaches are also employed, where the correlations of interest between two covariates relate to their means or percentages within groups. The primary difference with our study design is that we have access to the individual marginal observations of each variable. This allows us to conduct statistical inferences that ecological regression does not permit \citep{Gelman2001a}.

In this work, we assume that there is a linear relationship between the continuous variables that are not simultaneously observed, and that the linear regression coefficients are the same for all sub-populations defined by the categorical variable. A similar context is treated in \cite{evans_doubly_2021}, where a more general model is considered. The authors propose an approach that requires correctly specifying various relationships between variables (e.g., the distribution of the predictor variable conditional on what corresponds to the grouping variable in our case) in order to perform successful inference in the general case. Their approach is illustrated using survey data, which is the case when the latter can be expected to be successful due to large samples and/or prior knowledge. However, this is not the case for the experimental data considered in this work, which are characterized by small sample sizes and lack of prior knowledge about the underlying distributions. To address the specificities of the considered context, our approach is based on the assumption that there is a linear relationship conditional on the group, and that this relationship is the same for all groups, without making any assumptions about the distributions of the variables or requiring their specification.

We propose two estimators derived with the method of moments as well as an optimal transport solution using Wasserstein distance. Both approaches do not require any overlap between the two cohorts of the experiment and are based on weak assumptions ensuring model identifiability and the existence of moments. 
 These  estimators  are  shown  to  be  consistent  and  asymptotically  Gaussian  under weak hypotheses.  The asymptotic variance has no explicit expression, except in some special simple cases. For that reason a consistent stratified bootstrap approach is developed to construct confidence intervals for the estimated parameters. Not previously considered in the missing data literature, the bootstrap-based approach can be seen as the most important contribution of this paper, since the asymptotic solution is often infeasible in practice, whereas the bootstrap shows practical advantages in many realistic settings, especially in the case of a small sample size.

\section{Identification approaches}
We consider a real random variable $Y$ and a vector of $d$ real valued random regressors $\mathbf{X} = (X_1, \ldots, X_d)$, and suppose that the following linear regression holds:
\begin{align}
	Y & = \beta_0 + \sum_{j=1}^d \beta_j X_j + \epsilon.
	\label{def:lmm}
\end{align}
The residuals $\epsilon$ are supposed to be independent of the random covariates  $X_1, \ldots, X_d$, with zero mean and variance $\sigma^2_\epsilon$.  
In in vivo experiments, conducted under the design depicted in Figure \ref{fig:vivo_exp_ex}, we do not observe $\mathbf{X}$ and $Y$ simultaneously, i.e. we do not have the pair $(\mathbf{X},Y)$ at hand, but only $(\mathbf{X},.)$ and $(.,Y)$. This means that only the marginal distributions of $\mathbf{X}$ and $Y$ can be estimated in the presence of sampled data.

In the absence of additional information and without a strong additional hypothesis, the parameters $(\beta_0,\beta_1, \ldots, \beta_d)$ and the variance of the noise $\sigma^2_\epsilon$ cannot be identified. For example, if $X_1$ is centered and has a symmetric distribution, the coefficient $\beta_1$ can only be determined up to the sign change, since $\beta_1 X_1$ and $\beta_1 (-X_1)$ have the same distribution.

To deal with this identification problem, we consider that we can perform different experiments in which the mean of $X$ is allowed to vary. To do this, we assume that there are $K$ groups (i.e. $K$ different experiments), defined by a %discrete or 
categorical variable $G$ taking values in $\{1, \ldots, K\} $ observed simultaneously with $Y$ and $X$. This means that we now have access to $(\mathbf{X},G)$ and $(Y,G)$, but not to $(\mathbf{X},Y,G)$. We also assume that $\epsilon$ is independent of $G$.

Given $G=k$, for $k =1, \ldots, K$, we denote by $\mu_Y^k = \E (Y | G=k)$ and $\mu_{X_j}^k = \E (X_j | G=k)$, $j=1, \ldots, d$, the expected values within each group. We present two different approaches to identify the vector $\boldsymbol{\beta} = (\beta_0, \ldots, \beta_d)$ of unknown regression coefficients and the noise variance $\sigma_\epsilon^2$, taking into account the additional information provided by the discrete variable $G$.  

\subsection{Moment approach}
The first simple approach is based on the first moments identification. Taking the conditional expectation, given $G=k$, in (\ref{def:lmm}), we have for $k=1, \ldots, K$,
\begin{align}
	\mu_Y^k & = \beta_0 + \sum_{j=1}^d \beta_j \mu_{X_j}^k, 
	\label{def:lmmG}
\end{align}
since the residual term $\epsilon$ is assumed to satisfy $\mathbb{E}(\epsilon | G=k) = 0$ for $k=1, \ldots, K$. 

We denote by $\boldsymbol{\mu}_{1,X}$ the $K \times (d+1)$ design matrix, with the $k$th row equal to $(1, \boldsymbol{\mu}_{X}^{k\top})$ with $\boldsymbol{\mu}_{X}^k=(\mu_{X_1}^k, \cdots,  \mu_{X_d}^k)^\top$, and by $\boldsymbol{\mu}_Y$ the $K$ dimensional vector with elements $(\mu_Y^1, \ldots, \mu_Y^K)$. The  $K$ linear equations in \eqref{def:lmmG} can be written in a matrix form: $\boldsymbol{\mu}_Y =  \boldsymbol{\mu}_{1,X} \boldsymbol{\beta}$.

The following assumption guarantees the identifiability of the model parameters:
\begin{align*}
	\mathbf{H}_1 & \quad \mbox{rank}(\boldsymbol{\mu}_{1,X}) =  d+1,
\end{align*}
meaning that there are at least $K \geq d+1$ groups and that the $d+1$ column vectors of $\boldsymbol{\mu}_{1,X}$ span a vector space of dimension $d+1$ in $\mathbb{R}^K$.

\begin{Lem}
	\label{lem:mm_ident}
	If the model \eqref{def:lmm} holds and the assumption $\mathbf{H}_1$ is fulfilled, $\boldsymbol{\beta}$ is uniquely identified in terms of the conditional first order moments of $\mathbf{X}$ and $Y$ given $G$,
	\begin{align*}
		\boldsymbol{\beta} &= \left( \boldsymbol{\mu}_{1,X}^\top \boldsymbol{\mu}_{1,X} \right)^{-1} \boldsymbol{\mu}_{1,X}^\top \boldsymbol{\mu}_Y.
		%			\label{eq:lmm_beta}
	\end{align*}
	Additionally, the noise variance $\sigma_\epsilon^2$ satisfies
	\begin{align*}
		\sigma_\epsilon^2 &= \sigma_Y^2 - \boldsymbol{\beta}_{-0}^\top \boldsymbol{\Gamma}_X  \boldsymbol{\beta}_{-0},
		%	\label{eq:lmm_var}
	\end{align*}
	where   $\sigma_Y^2$ is the variance of $Y$, $\boldsymbol{\Gamma}_X$ is the covariance matrix of $\mathbf{X}$ with elements $\mbox{Cov}(X_i, X_j) =  \sum_{k=1}^K \mbox{Cov}(X_i, X_j | G=k) \mathbb{P}[G=k]$ for $i$ and $j$ in $\{1, \cdots, d\}$, and $\boldsymbol{\beta}_{-0} = (\beta_1, \ldots, \beta_d)$.
\end{Lem}
The proof of Lemma \ref{lem:mm_ident} is direct and thus omitted. 

\subsection{Optimal transport approach}
The second approach is based on optimal transport, in particular on the idea of estimating the linear transformation of the distribution of $\mathbf{X}$ that is the closest to that of $Y$ with respect to the Wasserstein distance (see \cite{MR3939527} for a general introduction for statisticians). The optimal transport map T between Gaussian measures on $\mathbb{R}^d$ is linear, and the Wasserstein distance of order 2 between two Gaussian distributions $D_1$ and $D_2$, with $D_1 \sim \mathcal{N}(\boldsymbol{\mu}_{1}, \boldsymbol{\Gamma}_1)$ and $D_2 \sim \mathcal{N}(\boldsymbol{\mu}_{2}, \boldsymbol{\Gamma}_2)$, is equal to
\begin{align*}
	W_2^2(D_1,D_2) &=  \| \boldsymbol{\mu}_{1} - \boldsymbol{\mu}_{2} \|^2 + \mbox{tr} \left(  \boldsymbol{\Gamma}_1 + \boldsymbol{\Gamma}_2 - 2 \left( \boldsymbol{\Gamma}_2^{1/2} \boldsymbol{\Gamma}_1 \boldsymbol{\Gamma}_2^{1/2} \right)^{1/2} \right),
\end{align*}
where $\|.\|$ denotes  the Euclidean norm and $\mbox{tr}(\mathbf{A})$ the trace of  matrix $\mathbf{A}$.

If the linear model \eqref{def:lmm} holds, and if we assume that, given $G=k$,  $\mathbf{X}$ is a Gaussian random vector and $\epsilon$ is Gaussian, we have that $Y$ is also Gaussian given $G=k$, with expectation $\mu_Y^k = \beta_0 + \sum_{j=1}^d \beta_j \mu_{X_j}^k$ and variance $\sigma_\epsilon^2 + \boldsymbol{\beta}_{-0}^\top \boldsymbol{\Gamma}_X^k  \boldsymbol{\beta}_{-0}$, where   $\boldsymbol{\Gamma}_X^k$ is the variance matrix of $\mathbf{X}$ given $G=k$.
Thus, the Wasserstein distance between $D_\gamma$, the distribution of $\gamma_0 + \boldsymbol{\gamma}_{-0}^\top \mathbf{X}+ \epsilon$, and $D_Y$, the distribution of $Y$, is equal to 
\begin{align*}
	W_2^2 (D_{\gamma},D_{Y}) & =  \mathbb{E} \left[ W_2^2 (D_{\gamma},D_{Y}) | G \right] \\
	&= \sum_{k=1}^K \pi_k  \left[ (\mu_{Y}^k - \alpha_0 - \boldsymbol{\gamma}_{-0}^\top \boldsymbol{\mu}_{X}^k)^2 +  \left(\sigma_{Y,k} - \sqrt{\boldsymbol{\gamma}_{-0}^\top \boldsymbol{\Gamma}_X^k  \boldsymbol{\gamma}_{-0}+ \sigma_\epsilon^2} \right)^2 \right],
\end{align*}
where $\pi_k = \mathbb{P}[G=k]$ and $\sigma_{Y,k}^2 = \mbox{Var}(Y | G=k)$.

We introduce  the following loss criterion
\begin{align}
	\varphi(\boldsymbol{\gamma},\sigma^2) &= \sum_{k=1}^K \pi_k \left[ 
	(\mu_{Y}^k - \gamma_0 - \boldsymbol{\gamma}_{-0}^\top \boldsymbol{\mu}_{X}^k)^2 +  \left(\sigma_{Y,k} - \sqrt{\boldsymbol{\gamma}_{-0}^\top \boldsymbol{\Gamma}_X^k  \boldsymbol{\gamma}_{-0}+ \sigma^2} \right)^2
	\right].
	\label{eq:wass_g}
\end{align}
which evaluates the weighted Wasserstein distance between $Y$ and a linear combination of the $\mathbf{X}$ variables, contaminated by Gaussian noise, with variance $\sigma^2$.
We  state, without proof, the following lemma which ensures that the parameters $\boldsymbol{\beta}$ and $\sigma_\epsilon^2$ can be identified under general conditions:
\begin{Lem}
	\label{lem:ot_ident}
	If the model (\ref{def:lmm}) holds and the assumption $\mathbf{H}_1$ is fulfilled, $\varphi(\boldsymbol{\gamma},\sigma^2)$ has its unique minimum at  $\boldsymbol{\gamma} =\boldsymbol{\beta}$ and $\sigma^2 = \sigma_\epsilon^2$.
\end{Lem}

\section{Sampled data and estimators}
We assume that the experiments are performed for $K \geq 2$ different groups, and that for each group $k$, for $k=1, \ldots, K$ we have two independent samples $(Y^k_1, \ldots, Y^k_{n_y^k})$ and  $(X^k_{j,1}, \ldots, X^k_{j,n_x^k})_{j=1, \ldots, d}$, with sizes $n_y^k$ and $n_x^k$. For each unit  $i=1, \ldots, n_x^k$ from group $k$, the vector of   covariates is denoted by $\mathbf{X}_i^k = (X_{1,i}, \ldots, X_{d,i})$. We also define  $N_x = \sum_{k=1}^K n_x^k$ and $N_y = \sum_{k=1}^K n_y^k$, the total number of observations of the response $Y$ and the covariates $X_1, \ldots, X_d$.

As shown in Lemma \ref{lem:mm_ident} and Lemma \ref{lem:ot_ident}, the identification of the parameter $\boldsymbol{\beta}$ depends on the knowledge of the first two conditional moments, given $G$, of $Y$ and $\mathbf{X}$.
For $k=1, \ldots, K$, we denote by $\widehat{\mu}_{Y}^k = \frac{1}{n_y^k} \sum_{i=1}^{n_y^k} Y^k_i$ and $\widehat{\mu}_{X_j}^k = \frac{1}{n_x^k} \sum_{i=1}^{n_x^k} X^k_{j,i}$ the empirical mean within each group $k$ and  by
$\widehat{\sigma}^2_{Y,k}  = \frac{1}{n_y^k}  \sum_{i=1}^{n_y^k}  \left(Y_i^k\right)^2 - \left( \widehat{\mu}_{Y}^k  \right)^2$, and 
$\widehat{\boldsymbol{\Gamma}}^k_{X} = \frac{1}{n_x^k} \sum_{i=1}^{n_x^k}  \mathbf{X}_{i}^k (\mathbf{X}_{i}^k)^\top - \widehat{\boldsymbol{\mu}}_{X}^k (\widehat{\boldsymbol{\mu}}_{X}^k)^\top$, the empirical variances, 
with $\widehat{\boldsymbol{\mu}}_{X}^k = (\widehat{\mu}_{X_1}^k, \ldots, \widehat{\mu}_{X_d}^k)$.
We also define the total empirical mean and variance $\widehat{\mu}_{Y}  = \frac{1}{N_y} \sum_{k=1}^K {n_y^k} \widehat{\mu}_{Y}^k$, 
$\widehat{\boldsymbol{\mu}}_{X}  = \frac{1}{N_x} \sum_{k=1}^{K} {n_x^k} \widehat{\boldsymbol{\mu}}_{X}^k$, 
$\widehat{\sigma}^2_{Y}  = \frac{1}{N_y} \sum_{k=1}^K \sum_{i=1}^{n_y^k}  \left(Y_i^k\right)^2 - \left( \widehat{\mu}_{Y}  \right)^2$, 
$\widehat{\boldsymbol{\Gamma}}_{X}  = \frac{1}{N_x} \sum_{k=1}^K \sum_{i=1}^{n_x^k}  \mathbf{X}_{i}^k (\mathbf{X}_{i}^k)^\top   - 
\widehat{\boldsymbol{\mu}}_{X} \widehat{\boldsymbol{\mu}}_{X}^\top$. We denote by $\widehat{\boldsymbol{\mu}}_{1,X}$ the $K \times (d+1)$ matrix, with the first column consisting of ones, and the rest equal to $\widehat{\boldsymbol{\mu}}_{X}$. 

\subsection{Moment estimators}
If $\widehat{\boldsymbol{\mu}}_{1,X}$ is full rank, moment estimators of $\boldsymbol{\beta} = (\beta_0, \beta_1, \ldots, \beta_d)$ can be built by considering the empirical counterpart of the identification equations given in Lemma~\ref{lem:mm_ident},  
\begin{align}
	\widehat{\boldsymbol{\beta}}^M & = \left( \widehat{\boldsymbol{\mu}}_{1,X}^\top \widehat{\boldsymbol{\mu}}_{1,X} \right)^{-1} \widehat{\boldsymbol{\mu}}_{1,X}^\top \widehat{\boldsymbol{\mu}}_Y 
	\label{def:estmom_no_w}
\end{align}
where $\widehat{\boldsymbol{\mu}}_Y=(\widehat{\mu}_{Y}^1, \ldots, \widehat{\mu}_{Y}^K)$.
In the following, we consider a slightly more general moment estimator of $\boldsymbol{\beta}$, by introducing a weight $w_k$  given to each group $k$ of observations, with $w_k>0$ and $\sum_{k=1}^K w_k =1$. The weighted moment estimator $\boldsymbol{\beta}$ is defined as the minimizer of
\begin{align*}
	\psi(\boldsymbol{\gamma}) &= \sum_{k=1}^K w_k \left[ \widehat{\mu}_Y^k - \left(\gamma_0 + \sum_{j=1}^d \gamma_j \widehat{\mu}_{X_j}^k\right) \right]^2,
\end{align*} 
which is unique if $\widehat{\boldsymbol{\mu}}_{1,X}$ is full rank and defined by 
\begin{align}
	\widehat{\boldsymbol{\beta}}^M & = \left( \widehat{\boldsymbol{\mu}}_{1,X}^\top \mathbf{w} \widehat{\boldsymbol{\mu}}_{1,X} \right)^{-1} \widehat{\boldsymbol{\mu}}_{1,X}^\top \mathbf{w} \widehat{\boldsymbol{\mu}}_Y 
	\label{def:estmom}
\end{align}
where $\mathbf{w}$ is a diagonal matrix with diagonal elements $w_1, \ldots, w_K$. The first moment estimator considered in \eqref{def:estmom_no_w}  corresponds to the case with equal weights $w_k = K^{-1}$.

We can then define the following estimator of the noise variance, 
\begin{align}
	\widehat{\sigma}_\epsilon^{2, M}   &=  \widehat{\sigma}_Y^2 - (\widehat{\boldsymbol{\beta}}_{-0}^M)^\top \widehat{\boldsymbol{\Gamma}}_X  \widehat{\boldsymbol{\beta}}_{-0}^M.
	\label{def:estmom0}
\end{align}

\subsection{Optimal transport estimators}
Estimators of $\boldsymbol{\beta}$ and $\sigma^2_{\epsilon}$ based on an optimal transport criterion are derived by minimizing the empirical version $\varphi_n(\boldsymbol{\gamma},\sigma^2)$ of a functional $\varphi(\boldsymbol{\gamma},\sigma^2)$ defined by 
\begin{align}
	\varphi_n(\boldsymbol{\gamma},\sigma^2) &= \sum_{k=1}^K \pi_k \left[ 
	(\widehat{\mu}_{Y}^k - \gamma_0 - \boldsymbol{\gamma}_{-0}^\top \widehat{\boldsymbol{\mu}}_{X}^k)^2 +  \left(\widehat{\sigma}_{Y,k} - \sqrt{\boldsymbol{\gamma}_{-0}^\top \widehat{\boldsymbol{\Gamma}}_X^k  \boldsymbol{\gamma}_{-0}+ \sigma^2} \right)^2
	\right].
	\label{eq:wass_gn}
\end{align}
Note that in absence of a priori information on the probability $\pi_k$ of observing group $k$, we can set $\pi_k = K^{-1}$.
We denote by $(\widehat{\boldsymbol{\beta}}^W, \widehat{\sigma}^{2,W})$  the minimizers of $\varphi_n(\boldsymbol{\gamma},\sigma^2)$, which are obtained with iterative optimization algorithms based on gradient descent.
The algorithm can be initialized  with $(\widehat{\boldsymbol{\beta}}^M, \widehat{\sigma}^{2,M})$ .

\section{Consistency and asymptotic distribution}\label{sec:cons}
To study the asymptotic behavior of the estimators of $\boldsymbol{\beta}$ defined in the previous section, we assume that the number $K$ of groups is kept fixed, and that for all groups and all variables $\mathbf{X}$ and $Y$, the number of observations tends to infinity. This means that  $n_{\min} = \min( n_y^1, \cdots, n_y^K, n_x^1, \cdots,n_x^K)$, the smallest sample size among all experiments, should also tend to infinity. 

\begin{Lem}\label{lem:mm_consist}
	If   $\E(Y^2) < + \infty$ and $\E(\| \mathbf{X}\| ^2) < + \infty$, and the assumption $\mathbf{H}_1$ is fulfilled,  the sequence of
	estimators $(\widehat{\boldsymbol{\beta}}^M,\widehat{\sigma}_\epsilon^{2,M})$ defined by \eqref{def:estmom} and \eqref{def:estmom0} converges in probability to $(\boldsymbol{\beta}, \sigma_\epsilon^2)$ when $n_{\min}$ tends to infinity.
\end{Lem}

For the Wasserstein minimum distance estimator, since there is no explicit expression for the estimators, a compactness assumption is also made to obtain the consistency. 
\begin{Lem}\label{lem:ot_consist}
	If   $\E(Y^2) < + \infty$ and $\E(\| \mathbf{X}\| ^2) < + \infty$, 
	$(\boldsymbol{\beta}, \sigma^2_{\epsilon}) \in \Theta$ and $\Theta$ is a compact set that does not contain 0, 
	if the model (\ref{def:lmm}) holds and the hypothesis $\mathbf{H}_1$ is fulfilled, then the sequence of estimators $(\widehat{\boldsymbol{\beta}}^W,\widehat{\sigma}^{2,W}_{\epsilon})$ that minimize \eqref{eq:wass_gn} converges in probability to $(\beta, \sigma^2_{\epsilon})$ when $n_{\min}$ tends to infinity.
\end{Lem}

As far as the asymptotic distribution of the estimators is concerned, and for the sake of simplicity and simpler notation, from now on we will assume that the number of experiments is the same for all groups and all variables, i.e.  $n = n_y^1 = \ldots = n_y^K = n_x^1 = \ldots = n_x^K$.
% and $\pi_k = 1/K$, for $k=1, \ldots, K$. 
\begin{Prop}\label{prop:mm_asympt_norm}
	If the assumptions of Lemma \ref{lem:mm_consist} are fulfilled, as $n$ tends to infinity,
	\[
	\sqrt{n} \left( \widehat{\boldsymbol{\beta}}^M - \boldsymbol{\beta} \right) \rightsquigarrow \mathcal{N}\left( 0, \boldsymbol{\Gamma}_{\beta_M} \right)
	\]
	where the expression of the asymptotic covariance matrix $\boldsymbol{\Gamma}_{\beta_M}$ is given in the proof.
\end{Prop}
This result is based on the central limit theorem for empirical means and the application of the delta method, which involves computing the Jacobian of the inverse of matrices, making it difficult to obtain the explicit expression for $\boldsymbol{\Gamma}_{\beta_M}$  when $d > 1$. 

\begin{Rem}
	The weak convergence toward a Gaussian distribution presented in Proposition~\ref{prop:mm_asympt_norm} remains true, at the expense of heavier notation and a different asymptotic covariance matrix $\boldsymbol{\Gamma}_{\beta_M}$, provided that there exist two constants, $0< c \leq C$ such that 
	\begin{equation}
		0< c  \leq \frac{\max(n_y^1, \cdots, n_y^K, n_x^1, \cdots,n_x^K)}{\min(n_y^1, \cdots, n_y^K, n_x^1, \cdots,n_x^K)} \leq C < + \infty,
		\label{hyp:sampling_rate}
	\end{equation}
	and $\min(n_y^1, \cdots, n_y^K, n_x^1, \cdots,n_x^K) \to \infty$.
\end{Rem}

The asymptotic normality of $\widehat{\boldsymbol{\beta}}^{W}$ relies on classical results for M-estimators recalled in Supplementary Material (see Theorem \ref{the:newey_macfadden_3}). Note that since we estimate $\beta$ and $\sigma_\epsilon^2$ simultaneously, an additional condition on the existence of the moments of order four is required for the covariates.

\begin{Prop}
	\label{prop:ot_asympt_norm}
	If  the model (\ref{def:lmm}) holds, the hypothesis $\mathbf{H}_1$ is fulfilled,  $\E(Y^2) < + \infty$ and $\E(\| \mathbf{X}\| ^4) < + \infty$, 
	$(\boldsymbol{\beta}, \sigma^2_{\epsilon}) \in \Theta$ and $\Theta$ is a compact set that does not contain $(0,0)$, then, as $n$ tends to infinity,
	\begin{align*}
		\sqrt{n} \left( \begin{pmatrix} \widehat{\boldsymbol{\beta}}^W \\  \widehat{\sigma}^{2,W}_{\epsilon} \end{pmatrix} - \begin{pmatrix}  \boldsymbol{\beta} \\ \sigma^{2}_{\epsilon} \end{pmatrix} \right) 
		&\rightsquigarrow \mathcal{N}\left( 0,   \boldsymbol{\Gamma}_{W}  \right),
	\end{align*}
	for some covariance matrix $\boldsymbol{\Gamma}_{W}$.
\end{Prop}
Similarly to $\boldsymbol{\Gamma}_{\beta_M}$, the expression of the asymptotic covariance matrix $\boldsymbol{\Gamma}_{W}$ is almost impossible to explicitly derive manually, with the exception of some particularly simple cases. 

\section{The particular case of simple linear regression}
To illustrate the difficulty, consider the case of a simple linear regression, i.e. $d=1$. The following linear model 
\begin{align*}
	Y &= \beta_0 + \beta_1 X + \epsilon
\end{align*}
is assumed to hold, and if there are two groups $k$ and $j$ such that $\mu_X^k  \neq  \mu_X^j$, the identification assumption $\mathbf{H}_1$ is fulfilled and $\beta_0$ and $\beta_1$ can be uniquely determined.
The moment estimators of $\beta_0$ and $\beta_1$ defined in \eqref{def:estmom} have simple expressions:
\begin{align}
	\widehat{\beta}_0 & = \widehat{\mu}_{Y,w} - \widehat{\beta}_1 \widehat{\mu}_{X,w} \\
	\widehat{\beta}_1 & 
	= \frac{ \sum_{k=1}^K w_k \widehat{\mu}_{X}^k \widehat{\mu}_{Y}^k - \widehat{\mu}_{X,w}\widehat{\mu}_{Y,w}}{\sum_{k=1}^K w_k \left( \widehat{\mu}_{X}^k \right)^2 - \widehat{\mu}_{X,w}^2 }
\end{align}
where $\widehat{\mu}_{Y,w} = \sum_{k=1}^K w_k  \widehat{\mu}_{Y}^k$ and$\widehat{\mu}_{X,w} = \sum_{k=1}^K w_k  \widehat{\mu}_{X}^k$.

We focus on the asymptotic variance of the estimator $\widehat{\beta}_1$ of the slope parameter $\beta_1$, which is often the parameter of interest.

\begin{Lem} 
	\label{lem:d=1}
	Suppose that the model \eqref{def:lmm} is true, with $d=1$ and $K \geq 2$. If there are two groups $k$ and $j$ such that $\mu_X^k  \neq  \mu_X^j$ the vector $(\beta_0,\beta_1)$ is identifiable and,  as $n$ tends to infinity,
	\begin{align*}
		\sqrt{n} \left( \widehat{\beta}_1 - \beta_1 \right)  &\rightsquigarrow \mathcal{N} (0,\sigma^2_{\beta_1}) 
	\end{align*}
	with 		
	\begin{align*}
		\sigma^2_{\beta_1} 
		& = \frac{1}{ (\mbox{Var}_w(X))^2} \sum_{k=1}^K w_k^2 \left( \beta_1^2\sigma^2_{X,k} + \sigma^2_{Y,k} \right) \left( \mu_X^k- \mu_{X,w}\right)^2 
	\end{align*}	
	and $\mu_{X,w} = \sum_{k=1}^K w_k  \mu_{X}^k$ and $\mbox{Var}_w(X) = \sum_{k=1}^K w_k  \left(\mu_{X}^k - \mu_{X,w} \right)^2$.
\end{Lem}

Lemma \ref{lem:d=1} shows that even in a very simple framework (only one covariate) the asymptotic variance of the estimator of the slope $\beta_1$ is quite complicated. It also reveals that minimizing the asymptotic variance $\widehat{\beta}_1$ with respect to the weights $w_k$ is not a simple task.

\section{Bootstrapping for confidence intervals}
Since, as noted in the previous section, it is complicated to explicitly  compute the asymptotic variance matrix of $\widehat{\boldsymbol{\beta}}^M$ and $\widehat{\boldsymbol{\beta}}^W$, we consider stratified bootstrap approaches in order to build confidence sets for $\boldsymbol{\beta}$. Our bootstrap procedure takes into account the independence between the different groups $k=1, \ldots, K$ as well as the independence of the inputs $(X_1^k, \ldots, X_d^k)$ and the output $Y^k$ within each group. More formally, given $G=k$, the joint probability measure $\mathbb{P}^k$ of $Y$ and $\mathbf{X}$ is a product measure of the marginal measures $\mathbb{P}^k = \mathbb{P}_Y^k \otimes \mathbb{P}_{\mathbf{X}}^k$.

Within each group $k$, we draw, with equal probability and with replacement, $n_y^k$ observations among $Y_1^k, \ldots, Y_{n_y^k}^k$.  We also draw independently, with equal probability and with replacement, $n_x^k$ observations among $\mathbf{X}_1^k, \ldots, \mathbf{X}_{n_x^k}^k$. We  denote by $\mu_Y^{k*}$ and by  $\boldsymbol{\mu}_X^{k*}$ the empirical means and by  $\sigma_{Y,k}^{2,*}$ and by  $\boldsymbol{\Gamma}_{X}^{k*}$ the empirical variances of $Y$ and $\mathbf{X}$ in these bootstrap samples.
Bootstrapped estimators  $\boldsymbol{\beta}^{M,*}$ and $\boldsymbol{\beta}^{W,*}$ of $\boldsymbol{\beta}$ can now be computed by replacing the empirical moments by the bootstrap moments in \eqref{def:estmom} and  \eqref{eq:wass_gn}. To build confidence sets for the components of  $\boldsymbol{\beta}$  based on this bootstrap procedure, the bootstrap percentile technique described in Chapter 4 of \cite{Shao_Tu_1995} can be applied. 

It can be noted that our estimators are smooth functions of the sample means, so classical bootstrap theory applies (see for example \cite{Shao_Tu_1995}, Chapter 3). For simplicity, we assume, as in Proposition \ref{prop:mm_asympt_norm}, that $n = n_y^1 = \ldots = n_y^K = n_x^1 = \ldots = n_x^K$.
Because of the experimental design considered, our global "empirical distribution" consists of products of marginal empirical distributions, the bootstrap for the means is almost surely consistent for the Kolmogorov metric, and with Theorem 3.1 in \cite{Shao_Tu_1995} the same result holds for the estimators of $\boldsymbol{\beta}$ considered in this work.  The application of Theorem 4.1 in \cite{Shao_Tu_1995} allows to conclude that the bootstrap percentile method gives consistent confidence bounds for each component of $\boldsymbol{\beta}$.

\begin{Prop}\label{prop:bootM}
	Suppose that  $\mathbb{E}(Y^2) < \infty$ and $\mathbb{E}\| \mathbf{X} \|^2 < \infty$ and the hypothesis $\mathbf{H}_1$ is fulfilled. Then as $n \to + \infty$, 
	the bootstrap estimator $\boldsymbol{\beta}^{M,*}$ is strongly consistent for $\boldsymbol{\beta}$ in the Kolmogorov metric.
	Consider a risk $\alpha \in (0,1)$, then for a given nominal level $1-\alpha$, for each component of $\boldsymbol{\beta}$, the bootstrap percentile approach provides a consistent confidence set. 
\end{Prop}

We can also state a similar result for the estimators minimizing the Wasserstein distance, under slightly more restrictive moment conditions and a compactness assumption. 
\begin{Prop}\label{prop:bootW}
	Suppose that  the assumptions of Proposition \ref{prop:ot_asympt_norm} are fulfilled. %$\mathbb{E}(Y^2) < \infty$ and $\mathbb{E}\| \mathbf{X} \|^4 < \infty$ and hypothesis $\mathbf{H}_1$ is fulfilled. 
	As $n \to + \infty$, 
	the bootstrap estimator $(\boldsymbol{\beta}^{W,*},  \sigma_\epsilon^{2,W,*}) $ is strongly consistent for $(\boldsymbol{\beta},  \sigma_\epsilon^2)$ in the Kolmogorov metric.
	Consider a risk $\alpha \in (0,1)$, then for a given nominal level $1-\alpha$, for each component of $\boldsymbol{\beta}$, the bootstrap percentile approach provides a consistent confidence set. 
\end{Prop}	

Propositions  \ref{prop:bootM} and \ref{prop:bootW} are very important for practical applications since they ensure that even if we are unable to compute the explicit expression of the asymptotic variance of our estimators, the asymptotic confidence intervals can still be constructed with a simple bootstrap procedure.

\section{Simulation study}\label{sec:vivo_sim}
\subsection{Simulation design}
We performed a series of simulations to evaluate the finite sample performance of the proposed approaches on data resembling in vivo data from real experiments on mice. The number of animals observed per group is chosen for simplicity to be $n=n_y^1 = \cdots = n_y^K = n_x^1 \cdots = n_x^K$, and thus the weights for each group are also assumed to be equal, i.e. $w_1 = \cdots = w_K = \frac{1}{K}$. For each animal $i \in \{1, \dots, n\}$ and each subpopulation $k \in \{1, \dots, K \}$, the predictor variable $X^k_i$ is Gaussian univariate: $X^k_i \sim \mathcal{N} (\mu_X^k, \sigma_{X}^2)$, where $\mu_X^k=9 + k$. We simulate the predicted variable independently as $Y^k_i=\beta_0 + \beta_1X'^k_i + \epsilon^k_i$, where $X'^k_i \sim X^k_i$ and $\epsilon^k_i \sim \mathcal{N} (0, \sigma_{\epsilon}^2)$, with regression parameters $\beta_0=1$ and $\beta_1=2$. By making $X^k_i$ and $X'^k_i$ independent, we recreate the situation where the predictor and predicted variables are not observed simultaneously. 

The variable sets $X=(X^k_i)_{1\leq i \leq n, 1\leq k\leq K}$ and $Y=(Y^k_i)_{1\leq i \leq n, 1\leq k\leq K}$ are simulated $N_{sim}$ times. For each simulation, $N_{boot}$ bootstrap samples of size $n$ are generated from $X^k=(X^k_i)_{1\leq k\leq K}$ and $Y^k=(Y^k_i)_{1\leq k\leq K}$ for each subpopulation $k \in \{1, \dots, K \}$ independently, then the moment estimators $\mu_X^{k*}$ and $\mu_Y^{k*}$ are computed. Finally, the bootstrap sample-based estimators $\boldsymbol{\beta}^{M,*}=(\beta_0^{M,*}, \beta_1^{M,*})$ and $\boldsymbol{\beta}^{W,*}=(\beta_0^{W,*}, \beta_1^{W,*})$ are computed. Based on the $N_{boot}$ estimates, we compute 95\% confidence intervals using the \texttt{quantile} function from the Python library NumPy. As a result, we obtain $N_{sim}$ confidence intervals for each regression parameter, which we use to calculate the following quantities of interest: the coverage rate of the intervals, their average amplitude, and the power, i.e. the proportion of intervals that do not contain 0. In addition to the bootstrap estimators, we also considered asymptotic confidence intervals in the case of the method of moments, obtained by plugging the estimated values of the moments into the expression for the limit distribution presented in Lemma \ref{lem:d=1}. In addition, we estimate the confidence intervals for the parameter estimators of the regression on the means per group with a naive method, assuming that the deviation of the parameter estimator from the true value divided by the standard error of the estimator follows a Student's t-distribution:
$$
\frac{\widehat{\beta}_j - \beta_j}{SE(\widehat{\beta_j})}  \rightsquigarrow t_{K-(d+1)} \text{ for } j \in \{0, \dots, d\}.
$$
Finally, we considered the case where the predictor and the predicted variable are observed simultaneously, i.e. $X^k_i$ and $Y^k_i$ are such that $Y^k_i=\beta_0 + \beta_1X^k_i + \epsilon^k_i$. In this case the parameters are estimated with the classical linear regression approach, and the confidence intervals are obtained with a Student's t-distribution.

\begin{figure}
	\centering
	\begin{subfigure}{.32\textwidth}
		\includegraphics[clip, trim=2cm 1cm 13cm 1cm, width=\textwidth]{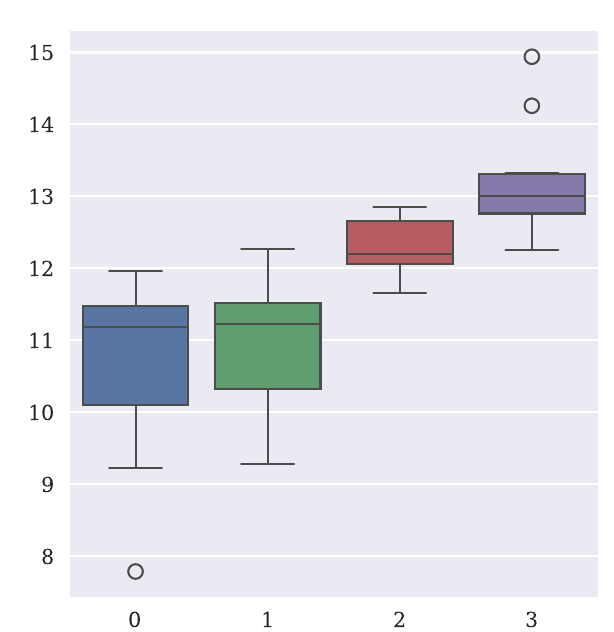}
		\caption{\label{fig:rho_x_box}}
	\end{subfigure}
	\begin{subfigure}{.32\textwidth}
		\centering
		\includegraphics[clip, trim=13cm 1cm 2cm 1cm, width=\textwidth]{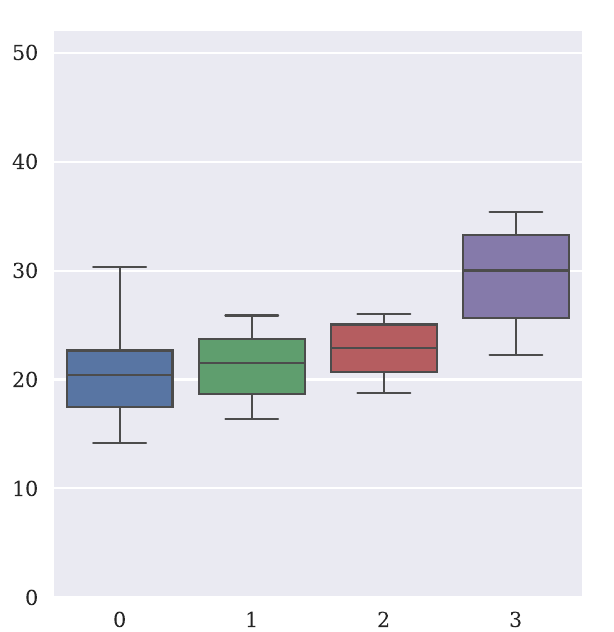}
		\caption{\label{fig:rho_11_y_box}}
	\end{subfigure}
	\begin{subfigure}{.32\textwidth}
		\centering
		\includegraphics[clip, trim=13cm 1cm 2cm 1cm, width=\textwidth]{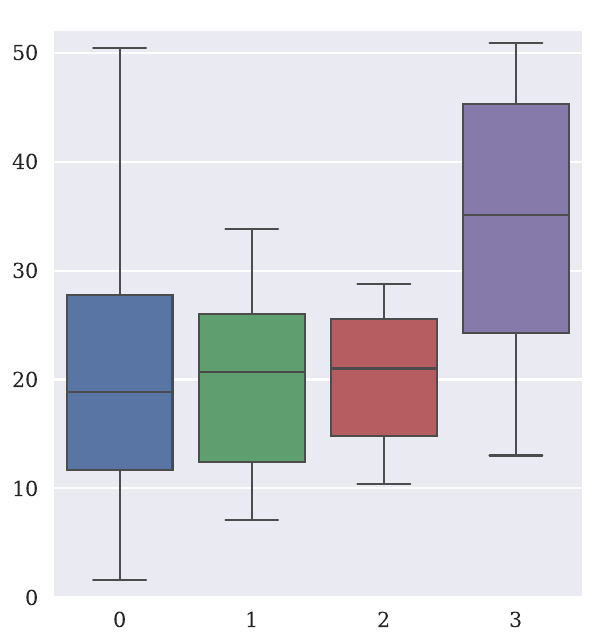}
		\caption{\label{fig:rho_101_y_box}}
	\end{subfigure}
	\caption{\label{fig:rho_effect}The effect of different values of $\rho$ on the data, with $K=4$ and $\sigma_{X}^2= 0.75$. a) Boxplots constructed from the simulated values of $X^k_i$. b) Boxplots constructed from the simulated values of $Y^k_i$ with lower relative noise level, i.e. $\rho=1.1$. c) Boxplots constructed from the simulated values of $Y^k_i$ with higher relative noise level, i.e. $\rho=1.01$.}
\end{figure}

Throughout all simulations we fix the number of simulations $N_{sim}=500$ and the number of bootstrap samples $N_{boot}=500$. Multiple parameters are varied to study their effect. We take the number of animals $n \in \{10, 30\}$, in particular to test whether inference is significantly affected when measurements are only available for a small number of animals, which is often the case in real experimental data. We consider the number of groups $K \in \{4, 10\}$, where 4 is the number of groups often observed in real data, and 10 is a higher number that may produce sufficiently good results with the naive approach of approximating confidence intervals with the Student distribution. Additionally, the parameter $\sigma^2_{X}$ can be adjusted to control the extent to which the observations per group can be easily distinguished from each other. We set $\sigma^2_{X} \in \{0.75, 2\}$, the first value corresponding to less overlap between groups and the second to more overlap. Finally, we introduce an additional parameter $\rho \in \R^{+}$ which controls the variance of the response to the variance of the noise ratio, i.e. $\rho=\frac{\sigma_{Y}}{\sigma_{\epsilon}}$, where $\sigma^2_{Y}=Var(Y^k_i)$ for all $i$ and $k$. The choice of adjusting the signal-to-noise ratio rather than the amount of noise itself through $\sigma_{\epsilon}^2$ is motivated by the fact that $\sigma_{Y}$ depends on $\sigma_{X}$, so the same level of $\sigma_{\epsilon}$ cannot be interpreted in the same way for different values of $\sigma_{X}$. The variance of the noise can be expressed as follows $\sigma_{\epsilon}^2=\frac{\beta_1^2 \sigma_{X}^2}{\rho^2 - 1}$. The values of $\rho$ are chosen to correspond to the realistic situation, namely a very noisy case and a slightly less noisy one: $\rho \in \{1.01, 1.1\}$. The effect of different values of $\rho$ on the simulated response variable is illustrated in Figure \ref{fig:rho_effect}.

\subsection{Results}\label{sec:vivo_sim_results}
\begin{figure}
	\centering
	%	\vspace{450pt}
	\begin{subfigure}{0.325\textwidth}
		\includegraphics[clip, trim=6cm 0.8cm 57cm 1.77cm,width=\textwidth]{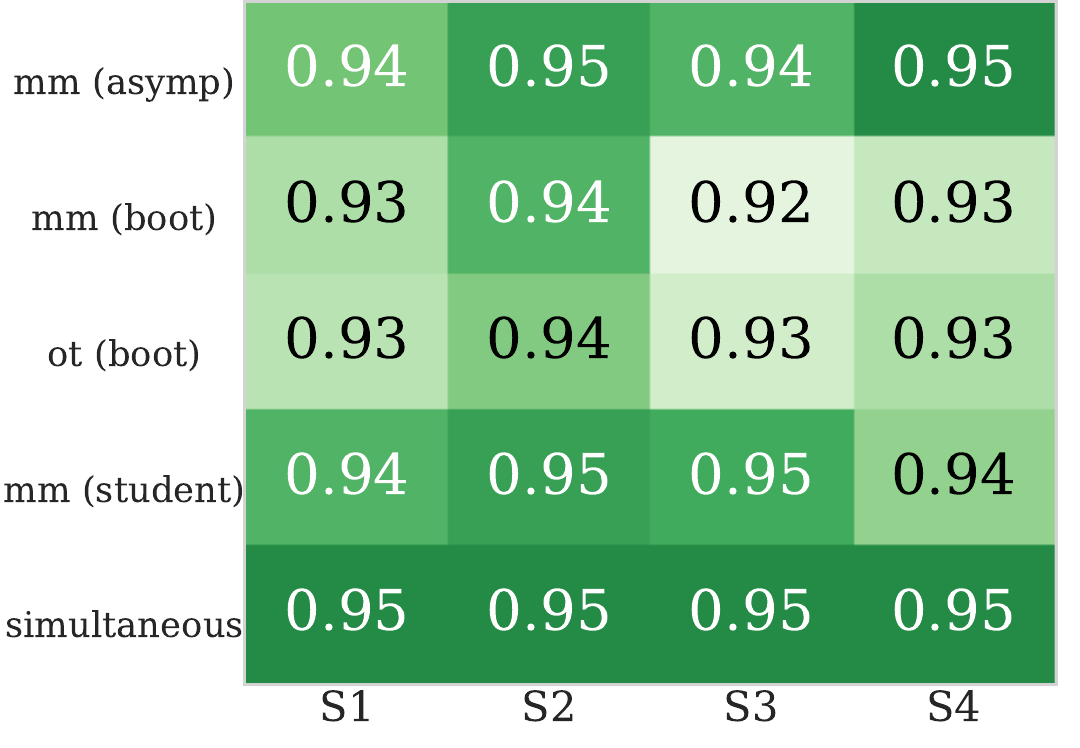}
		\caption{\label{fig:ci_cov_rate_410}}
	\end{subfigure}
	\begin{subfigure}{.325\textwidth}
		\includegraphics[clip, trim=6cm 0.8cm 57cm 1.77cm,width=\textwidth]{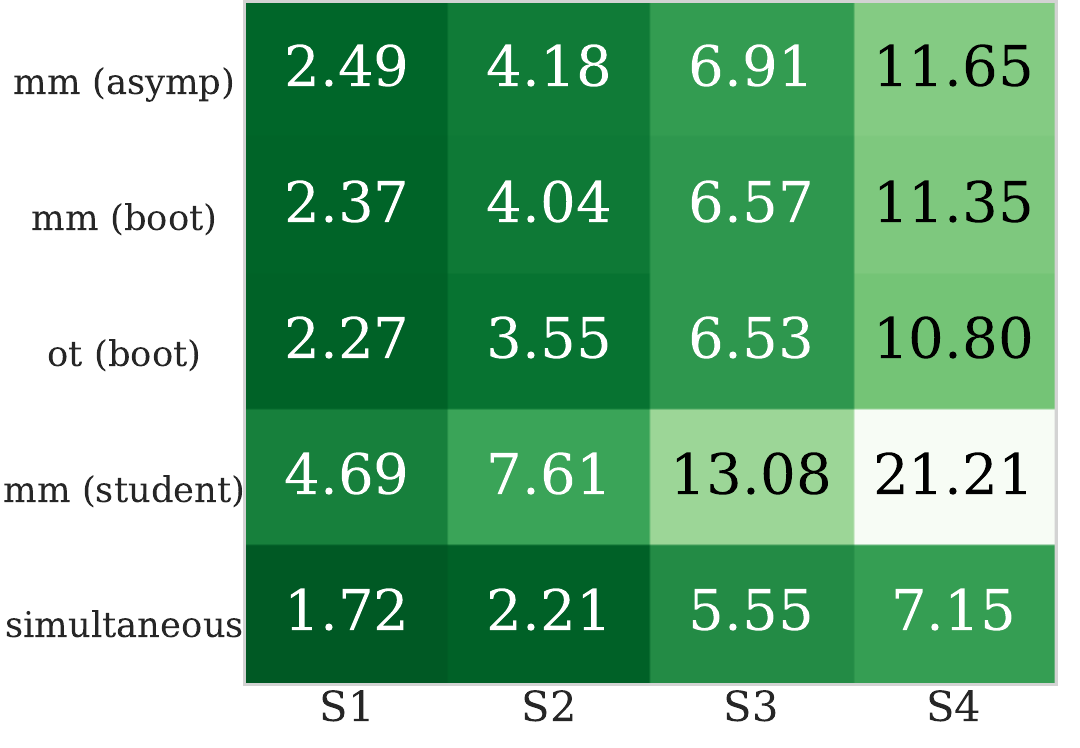}
		\caption{\label{fig:ci_amplitude_410}}
	\end{subfigure}
	\begin{subfigure}{.325\textwidth}
		\includegraphics[clip, trim=6cm 0.8cm 57cm 1.77cm,width=\textwidth]{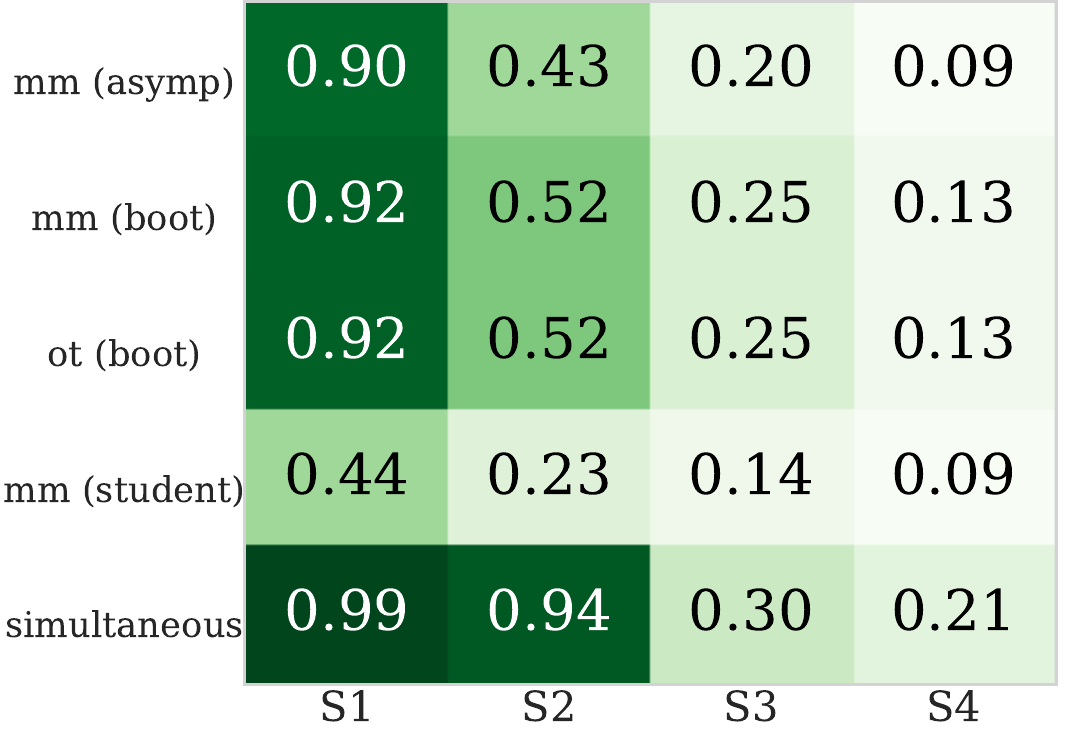}
		\caption{\label{fig:ci_power_410}}
	\end{subfigure}
	\caption{\label{fig:vivo_sim_results_410}a) Coverage rates, b) average amplitudes, and c) powers of the confidence intervals for the estimators of $\beta_1$ obtained from 500 simulations, with number of groups $K=4$ and number of animals per group $n=10$. The columns of the tables indicate simulation scenarios with different combinations of parameters: scenario S1 with lower group overlap ($\sigma^2_{X}=0.75$) and higher signal-to-noise ratio ($\rho=1.1$), S2 with higher group overlap ($\sigma^2_{X}=2$) and higher signal-to-noise ratio ($\rho=1.1$), S3 with lower group overlap ($\sigma^2_{X}=0.75$) and lower signal-to-noise ratio ($\rho=1.01$), and S4 with higher group overlap ($\sigma^2_{X}=2$) and lower signal-to-noise ratio ($\rho=1.01$). The lines indicate the method used to estimate the confidence intervals: "mm (asymp)" stands for the method of moments with asymptotic confidence intervals, "mm (boot)" for the method of moments with bootstrap, "ot (boot)" for the optimal transport method with bootstrap, "mm (student)" for the naive linear regression on means approach based on Student's distribution, and "simultaneous" for the classical linear regression estimation in the case where the predictor and the predicted variable are observed simultaneously.}
\end{figure}

The results of the simulation study are shown in Supplementary Figure \ref{fig:vivo_sim_results}. Firstly, the results for the moment approach are very similar in the asymptotic and bootstrap cases, confirming the effectiveness of the bootstrap approach. When comparing the bootstrap procedure with the naive approach, it can be observed that the bootstrap estimators generally produce confidence intervals with smaller average amplitudes and higher power at the expense of a slightly lower coverage rate. Whereas the empirical coverage rates are close to the nominal one (95\%) in all cases, the extent to which the average amplitudes are smaller and the powers are greater for the bootstrap estimators is very important in almost all cases. This implies that the naive approach based on the Student's distribution is more likely to produce false negatives in terms of significance. This trend is further strengthened by the parameter encoding the number of groups: while the overall results worsen with a decrease in either the number of animals or the number of groups, it is the case with a small number of groups that shows the greatest difference between the approaches (the case with few groups and animals is shown in Figure \ref{fig:vivo_sim_results_410}). Indeed, in almost all cases within the tables with $K=4$ we observe that the average amplitudes are approximately twice as important for the naive approach, and a similar trend in terms of lower powers. The latter result is important because lower power implies a higher probability of not detecting a significant relationship between the predictor and the predicted variables, if it is indeed present. Overall, these results imply that the proposed bootstrap estimators are more effective when the experimental design involves a small number of groups.

Regarding the remaining two parameters, as expected, the best results are generally obtained with lower $\sigma^2_{X}$ and higher $\rho$. In most cases, the results for the naive and bootstrap estimators are either both good or both bad in terms of power, with the latter being slightly better. A particularly complicated case can be distinguished, with high overlap, high noise, few groups and few animals, where all estimators fail drastically: we observe almost equally bad powers (0.1 for both bootstrap estimators and 0.09 for the naive estimator), despite the significant difference in average amplitudes. On the other hand, we can also distinguish two cases where the powers of the bootstrap estimators are above 90\% while those of the naive estimator are below 50\%: in both cases there are 4 groups and high noise, in the first case there are only 10 animals but less overlap, in the second case a high level of overlap is compensated by a higher number of animals. This means that if the underlying distributions per group are characterized by a reasonable amount of overlap, or if a significant overlap is compensated by having more observations, the bootstrap estimators manage to detect the significant relationship in most cases, unlike the naive estimator.

\begin{figure}
	\centering
	%	\vspace{450pt}
	\begin{subfigure}{0.47\textwidth}
		\includegraphics[clip, trim=1.5cm 1cm 1.5cm 1.5cm, width=\textwidth]{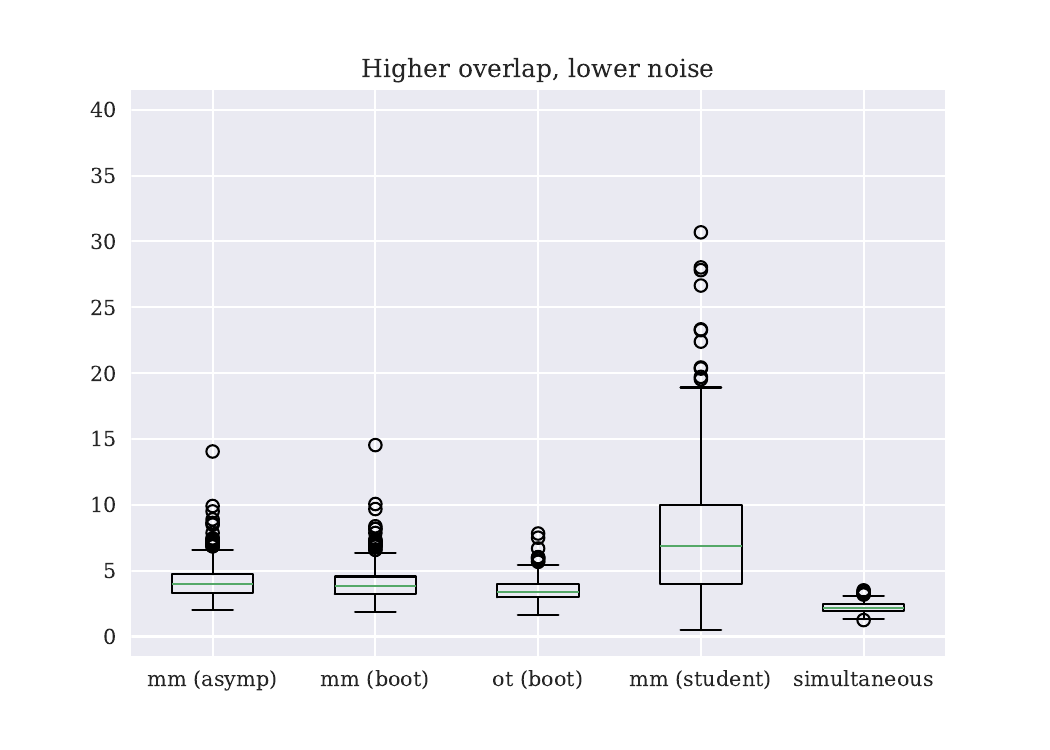}
		\caption{\label{fig:sim_box_2-11}}
	\end{subfigure}
	\begin{subfigure}{.47\textwidth}
		\includegraphics[clip, trim=1.5cm 1cm 1.5cm 1.5cm, width=\textwidth]{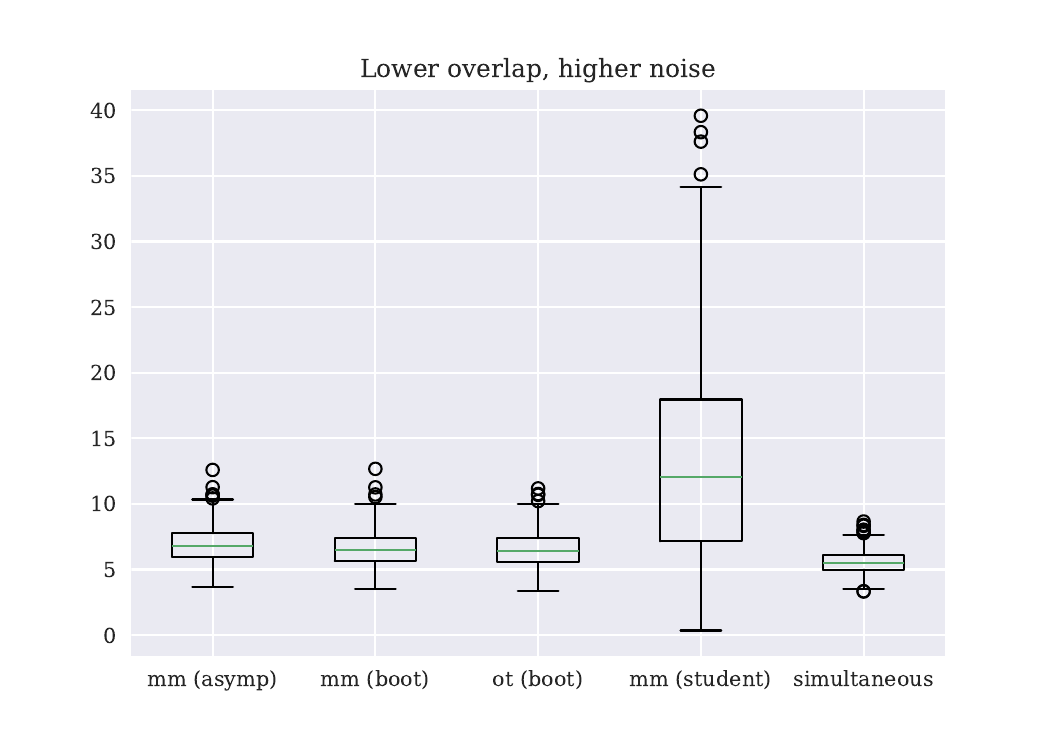}
		\caption{\label{fig:sim_box_075-101}}
	\end{subfigure}
	\caption{\label{fig:sim_boxplots}Distributions of amplitudes of confidence intervals obtained with different methods based on 500 simulations under scenarios S2 (a) and S3 (b), with $K=4$ and $n=10$.}
\end{figure}

Concerning the comparison with the simultaneous case, as expected, the latter globally produces better results than any method in the context of non-simultaneous design. In particular, average amplitudes are systematically 10\%-40\% smaller compared to our methods in the non-simultaneous case. However, the difference seems to be less important than that between our approaches and the naive approach. Boxplots of the amplitude distributions for the most challenging combination of $K$ and $n$ under the most interesting scenarios (S2 and S3) are shown in Figure \ref{fig:sim_boxplots}. The figure confirms the fact that the amplitude distributions obtained with our methods are more similar to those obtained in the simultaneous case than to those obtained with the naive approach in the non-simultaneous case. Moreover, the powers are almost as high as those obtained in the simultaneous case in 3 out of 4 scenarios. The only scenario where this difference is striking is the one with little noise and high group overlap. It can be concluded that if the groups are separable, in the case where the predictor and the predicted variable are not observed simultaneously, with our approach we can detect a significant relationship with a success comparable to the case when the variables are observed simultaneously. It should be noted that if the data are too noisy, all methods fail to detect the significance in all cases.

Finally, it can be observed that the estimator based on optimal transport produces confidence intervals with slightly smaller average amplitudes compared to the method of moments estimator. The difference appears to be relatively more important in cases with higher group overlap $\sigma^2_{X}=2$. However, the powers are not affected by this difference. This may be explained by a more important bias associated with the optimal transport estimator. The estimator is likely to produce better results in terms of power than the method of moments estimator when the bias is corrected. 

\section{Application to real data}\label{sec:vivo_real_data}
To illustrate the proposed estimators on real data, we examined the data mentioned in~Section \ref{sec:intro}, obtained from experiments conducted in mice to assess the adverse effects induced in the context of different irradiated volumes. In these experiments (see \cite{bertho_preclinical_2020} for a more detailed presentation), mice were exposed to either stereotactic body radiation therapy (SBRT) with different beam sizes at 90 Gy to the left lung, or whole thorax irradiation (WTI) at 19 Gy. For one cohort of mice, the expression of pro-inflammatory genes (IL6 and TNF) was measured, and for the other cohort, the thickness of the alveolar septum was measured as a measure of the severity of radiation-induced lung lesions. In the case of SBRT, measurements were taken at multiple sites: the irradiated patch (within the irradiation field), the remaining part of the left lung, referred to as the ipsilateral lung, and the right lung (contralateral lung). A control condition is also included where gene expression is measured without prior exposure to radiation. The goal of this statistical analysis is to determine whether there is a statistical association between gene expression as a predictor and septal thickening as an outcome. Our approach is applied because the variables are measured on different animals, but within each irradiation condition there are common groups in terms of measurement time points.

To ensure comparability of the results for different treatment conditions and genes, the data were centered and reduced with respect to the global mean and standard deviation prior to estimation. The linear regression parameters were then estimated with three estimators in the same way as in the simulation study in Section \ref{sec:vivo_sim}. The focus is on the estimation of the slope parameter $\beta_1$. The results are presented in Table \ref{tab:vivo_results}, which contains the estimates of $\beta_1$ as well as the estimated confidence intervals for the slope estimator and the corresponding test result for the significance of the estimated relationship.

\begin{table}
\caption{\label{tab:vivo_results}Results of estimation of the linear regression slope predicting septal thickening with the pro-inflammatory genes expression, with three methods, for control (no irradiation), WTI and SBRT with different beam sizes, with measurements taken in different parts of lungs.}

\addtolength{\tabcolsep}{-2pt}
\begin{spacing}{1.5}
	\begin{scriptsize}
		\begin{tabular}[c]{|c|c|c|c|c|c|c|c|c|c|c|c|  } 
			\cline{4-12}
			\multicolumn{3}{c|}{} & \multicolumn{3}{c|}{\textbf{\textsc{Method of moments}}} & \multicolumn{3}{c|}{\textbf{\textsc{Optimal transport}}}  & \multicolumn{3}{c|}{\textbf{\textsc{Lin. Reg. on means}}} \\ 
			\multicolumn{3}{c|}{} & \multicolumn{3}{c|}{\textbf{\textsc{(bootstrap)}}} & \multicolumn{3}{c|}{\textbf{\textsc{(bootstrap)}}}  & \multicolumn{3}{c|}{\textbf{\textsc{(Student)}}} \\ 
			 \hline
			\textbf{\textsc{Loc.}} & \textbf{\textsc{Vol.}} & \textbf{\textsc{Gene}} & \textbf{\textsc{${\widehat{\beta}_1}$}} & \textbf{\textsc{95\% C.I.}} & \textbf{\textsc{Signif.}} & \textbf{\textsc{${\widehat{\beta}_1}$}} & \textbf{\textsc{95\% C.I.}} & \textbf{\textsc{Signif.}} & \textbf{\textsc{${\widehat{\beta}_1}$}} & \textbf{\textsc{95\% C.I.}} & \textbf{\textsc{Signif.}} \\ 
			\hline
			\multicolumn{2}{|c|}{\textbf{\textsc{Control}}} & IL6   & 2.23 & (-1.99, 2.28) & \textcolor{red}{\ding{55}} & 0.83 & (-0.83, 0.97) & \textcolor{red}{\ding{55}} & 2.23 & (-2.65, 7.12) & \textcolor{red}{\ding{55}} \\ 
			\cline{3-12}
			\multicolumn{2}{|c|}{}  & TNF &  2 & (-2.0, 2.82) & \textcolor{red}{\ding{55}} & 0.88 & (-1.0, 1.11) & \textcolor{red}{\ding{55}} & 2 & (-1.72, 5.72) & \textcolor{red}{\ding{55}} \\ 
			\hline
			& 1 mm & IL6   & 0.43 & (-0.23, 1.2) & \textcolor{red}{\ding{55}} & 0.2 & (-0.2, 0.84) & \textcolor{red}{\ding{55}} & 0.43 & (-1.32, 2.17) & \textcolor{red}{\ding{55}} \\ 
			\cline{3-12}
			\textbf{\textsc{Ipsilateral}}  & & TNF   & 0.2 & (-0.23, 0.84) & \textcolor{red}{\ding{55}} & 0.16 & (-0.18, 0.84) & \textcolor{red}{\ding{55}} & 0.2 & (-1.26, 1.66) & \textcolor{red}{\ding{55}} \\ 
			\cline{2-12}
			\textbf{\textsc{lung}}  & 3 mm  & IL6   & 0.05 & (-0.34, 0.46) & \textcolor{red}{\ding{55}} & 0.06 & (-0.33, 0.49) & \textcolor{red}{\ding{55}} & 0.05 & (-1.65, 1.76) & \textcolor{red}{\ding{55}} \\ 
			\cline{3-12}
			&  & TNF   & 0.65 & (-0.12, 1.6) & \textcolor{red}{\ding{55}} & 0.63 & (-0.12, 1.46) & \textcolor{red}{\ding{55}} & 0.65 & (-1.57, 2.87) & \textcolor{red}{\ding{55}} \\ 
			\hline
			& 1 mm  & IL6   & 1.03 & (-0.57, 2.19) & \textcolor{red}{\ding{55}} & 0.46 & (-0.3, 1.0) & \textcolor{red}{\ding{55}} & 1.03 & (-0.88, 2.94) & \textcolor{red}{\ding{55}} \\ 
			\cline{3-12}
			\textbf{\textsc{Right}} &  & TNF   & 1.05 & (-0.47, 2.43) & \textcolor{red}{\ding{55}} & 0.66 & (-0.31, 1.26) & \textcolor{red}{\ding{55}} & 1.05 & (-1.01, 3.11) & \textcolor{red}{\ding{55}} \\ 
			\cline{2-12}
			\textbf{\textsc{lung}}  & 3 mm  & IL6   & 0.3 & (-1.45, 1.12) & \textcolor{red}{\ding{55}} & 0.37 & (-0.74, 0.86) & \textcolor{red}{\ding{55}} & 0.3 & (-4.94, 5.53) & \textcolor{red}{\ding{55}} \\ 
			\cline{3-12}
			&  & TNF   & 2.02 & (0.27, 4.1) & \textcolor{green}{\ding{52}} & 1.05 & (0.04, 1.44) & \textcolor{green}{\ding{52}} & 2.02 & (0.03, 4.02) & \textcolor{green}{\ding{52}} \\ 
			\hline
			& 1 mm  & IL6   & 0.85 & (-0.7, 2.38) & \textcolor{red}{\ding{55}} & 0.61 & (-0.56, 1.6) & \textcolor{red}{\ding{55}} & 0.85 & (-3.75, 5.45) & \textcolor{red}{\ding{55}} \\ 
			\cline{3-12}
			\textbf{\textsc{Irradiated}} &  & TNF  & 0.85 & (-0.6, 2.3) & \textcolor{red}{\ding{55}} & 0.69 & (-0.54, 1.53) & \textcolor{red}{\ding{55}} & 0.85 & (-3.96, 5.66) & \textcolor{red}{\ding{55}} \\ 
			\cline{2-12}
			\textbf{\textsc{patch}}  & 3 mm  & IL6   & 1.35 & (0.22, 2.47) & \textcolor{green}{\ding{52}} & 1.3 & (0.21, 2.26) & \textcolor{green}{\ding{52}} & 1.35 & (-1.8, 4.5) & \textcolor{red}{\ding{55}} \\ 
			\cline{3-12}
			&  & TNF   & 3.81 & (1.01, 6.37) & \textcolor{green}{\ding{52}} & 3.37 & (0.99, 5.33) & \textcolor{green}{\ding{52}} & 3.81 & (-1.86, 9.48) & \textcolor{red}{\ding{55}} \\ 
			\hline
			\multicolumn{2}{|c|}{\textbf{\textsc{Whole thorax}}}  & IL6 & 3.7 & (1.53, 5.99) & \textcolor{green}{\ding{52}} & 2.51 & (1.39, 3.43) & \textcolor{green}{\ding{52}} & 3.7 & (1.11, 6.29) & \textcolor{green}{\ding{52}} \\ 
			\cline{3-12}
			\multicolumn{2}{|c|}{\textbf{\textsc{irradiation}}}  & TNF & 2.35 & (0.57, 4.73) & \textcolor{green}{\ding{52}} & 1.67 & (0.53, 1.88) & \textcolor{green}{\ding{52}} & 2.35 & (-3.86, 8.57) & \textcolor{red}{\ding{55}} \\ 
			\hline
		\end{tabular}
	\end{scriptsize}
\end{spacing}
\end{table}

Considerable differences can be observed between the results for the naive and the bootstrap estimators with respect to the significance found. On the one hand, the relationship between the pro-inflammatory genes and septal thickening was identified by all methods in the case of whole thorax irradiation (both genes for the bootstrap estimators and only one gene for the naive estimator), which is an expected result. The results are also consistent for all methods in the case of no radiation exposure (control), where no significant association was identified, as expected. On the other hand, we expect to identify a strong correlation in the case of measurements taken directly from the irradiated patch. This is the case only for the bootstrap estimators, but not for the naive one. This result is in accordance with the results obtained with simulated data: the confidence intervals are often overestimated with the naive approach, which can lead to false negatives in terms of significance. 

Among the SBRT irradiation configurations, only the 3 mm beam size showed a significant correlation between inflammatory genes and septal thickening. These results are consistent with the literature, indicating that this is the beam size from which the long-term lesions appear \citep{bertho_preclinical_2020}. Several significant associations were identified with the bootstrap estimators in the ipsilateral lung and in the right lung for the beam size of 3 mm.

\begin{figure}
	%\vspace{-0.5cm}
	\centering
	\includegraphics[clip, trim=0cm 0cm 0cm 1.5cm,width=\textwidth]{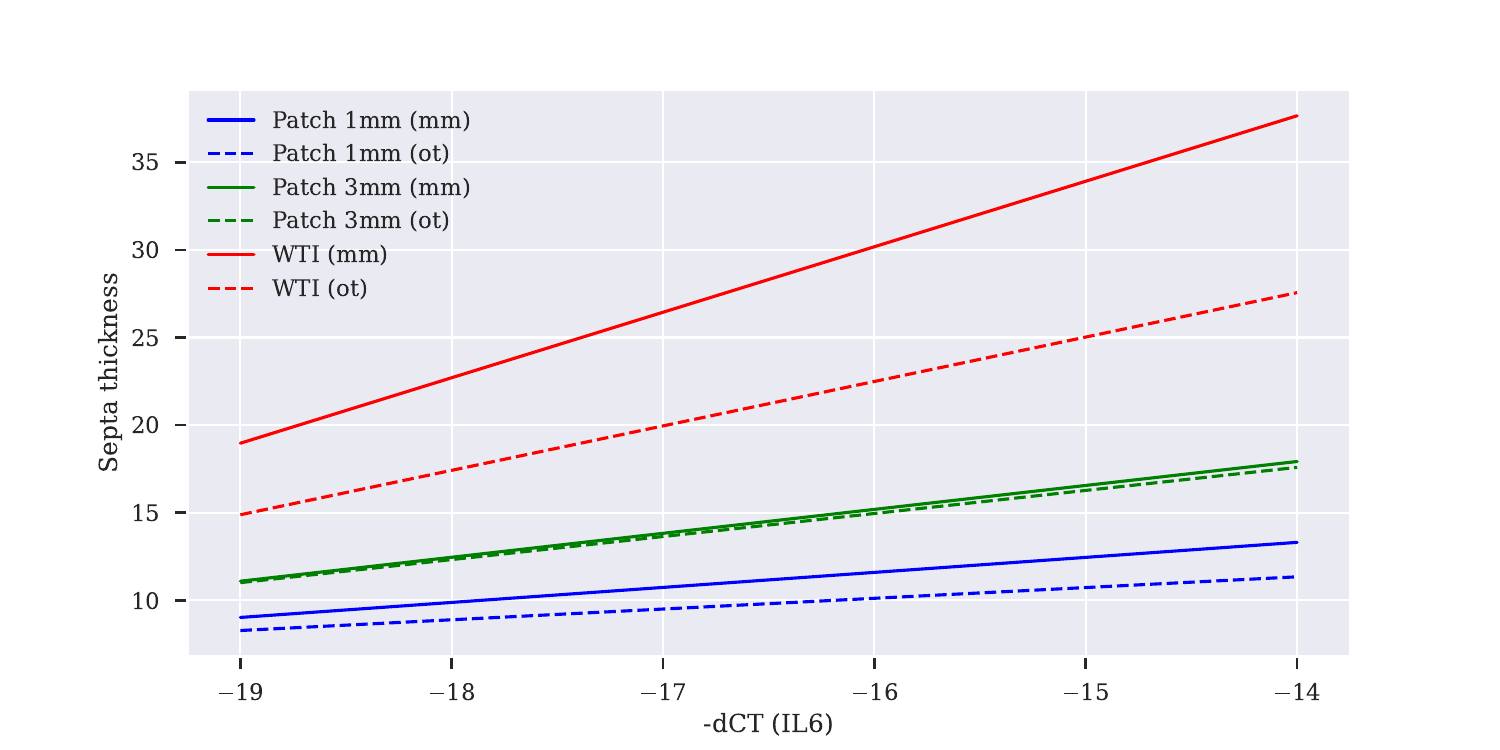}
	\caption{\label{fig:vivo_LR_predict}Linear model prediction of septal thickness based on IL6 expression, plotted for different locations and beam sizes, with the results from two bootstrap estimators.}
\end{figure}

Finally, in the cases where a significant relationship was detected, the estimated values of the slope are always positive, indicating a general radio-induced upregulation trend. These values are generally greater for whole thorax irradiation and within the patch than for the ipsilateral or right lung for SBRT, suggesting a stronger correlation between the inflammatory process and lung injury under high dose/volume irradiation conditions. These results, which are in line with biological knowledge, could not have been obtained using classical statistical regression approaches due to non-simultaneous observations. This effect is illustrated in Figure \ref{fig:vivo_LR_predict} using the example of linear model prediction for the gene IL6. 

\section{Discussion}\label{sec:discuss}
This work focuses on a statistical framework designed to extract dependencies from experiments, specifically introducing linear regression estimators in the context where the predictor and predicted variables are not jointly observed but share a common observed categorical variable. In this work we have chosen the basic linear multivariate setting, prioritizing simplicity and computational feasibility. In particular, the estimator based on the method of moments makes no hypotheses about the data distribution and can be computed explicitly. The optimal transport estimator involves a simple optimization problem and is based on the Gaussian form of the Wasserstein distance, but does not technically require the data to be Gaussian, seeking to approximate them with Gaussian variables in any case. The proposed bootstrap procedure produces confidence intervals for the regression parameters that are smaller than those obtained with the naive approach, while preserving a high coverage rate. In practice, this allows better detection of significant effects in cases where the sample size is small, which is often the case in in vivo experiments.

However, these approaches are not applicable in cases where the linear relationship hypothesis cannot be satisfied. For example, this is the case when predicting survival data with some continuous biomarkers, which is of particular interest in the research on the adverse effects of radiation. To be able to consider such scenarios, our model can be extended to a more general case, namely with a generalized linear model. The optimal transport estimator seems promising in this context, given the fact that the Wasserstein distance allows to compare probability distributions of different nature (e.g. continuous and discrete). 

Another potential direction of research is to investigate alternative methods based on integrated likelihood and Bayesian approaches, which are likely to produce better results in many cases, but require the imposition of priors on distributions.

Finally, it would be of interest to work on improving the theoretical properties of our finite sample estimators, namely the correction of the negative bias that appears for both estimators. The latter is particularly important in the case of the optimal transport estimator, which can arise naturally with the Wasserstein distance (e.g. \cite{manole2024plugin}). Correcting this bias would considerably improve the estimator, making it competitive with the aforementioned approaches, which make numerous assumptions about the data.

\section*{Acknowledgements}
The authors gratefully acknowledge the funding from the European Union through the PO FEDER-FSE Bourgogne 2014/2020 programs as part of the ModBioCan2020 project, and from the Institut de Radioprotection et de Sûreté nucléaire as part of the ROSIRIS project. The authors would also like to express their gratitude to Dr. Agnès François, Dr. Annaïg Bertho, Dr. Morgane Dos Santos and Dr. Fabien Milliat for the experimental data. Finally, the authors warmly thank Dr. Agnès François for her help in the biological interpretation of the data, as well as Dr. Patrick Tardivel for his numerous remarks that helped to improve the presentation of the manuscript. 

\section*{Supporting Information}
The Python code is freely available at \href{https://github.com/parsenteva/vivo\_lm}{github.com/parsenteva/vivo\_lm}. The data that support the findings in this paper are available upon request from Dr. Agnès François (agnes.françois@irsn.fr).

\bibliographystyle{apalike}
\bibliography{biblio_reg}

\newpage
\appendix

\section{Some classical theorems in asymptotic statistics}\label{sec:theorems}
A proof of the classical continuous mapping theorem can be found in \cite{MR1652247} (Theorem 2.3).

\begin{The} (Continuous mapping theorem). \\
	Let $g : \mathbb{R}^d \to \mathbb{R}^m$ be continuous at every point of $C$ such that $\mathbb{P}[ X \in C] = 1$. \\
	If the sequence of random variables $(X_n)_{n \geq 1}$ converges in distribution (resp. probability, resp. almost surely) to $X$ then $(g(X_n))_{n \geq 1}$ converges in distribution (resp. probability, resp. almost surely) to $g(X)$.
\end{The}

We also recall some well known results that are useful to show the consistency of estimators $\widehat{\theta}_n$ defined as the minimizers of functionals $Q_n(\theta)$ which have some regularity properties at the limit. 

\begin{The} (Lemma 2.9 in  \cite{MR1315971}) \\
	Suppose that $\theta \in \Theta$ and $\Theta$ is compact, $Q_0(\theta)$ is continuous and $\forall \theta \in \Theta$, $Q_n(\theta) \to Q_0(\theta)$ in probability as $n$ tends to infinity. If there is $\alpha >0$ and $B_n = O_p(1)$ such that 
	\[
	\forall (\widetilde{\theta}, \theta) \in \Theta \times \Theta, \ |  Q_n(\widetilde{\theta}) - Q_n(\theta) | \leq B_n \| \widetilde{\theta}  - \theta \|^\alpha
	\]
	then 
	\[
	\sup_{\theta \in \Theta} \left| Q_n(\theta) - Q_0(\theta)  \right| \to 0 \ \mbox{ in probability.} 
	\]
	\label{the:newey_macfadden_1}
\end{The}

\begin{The} (Theorem 2.1 in \cite{MR1315971}) \\
	Suppose that $\theta \in \Theta$ and $\Theta$ is compact, $Q_0(\theta)$ is continuous $\forall \theta \in \Theta$. If $Q_0(\theta)$ is uniquely maximized at $\theta_0$ and, as $n$ tends to infinity,
	$\sup_{\theta \in \Theta} \left| Q_n(\theta) - Q_0(\theta)  \right| \to 0$  in probability, then $\widehat{\theta}_n \to \theta_0$ in probability.   
	\label{the:newey_macfadden_2}
\end{The}

Under additional hypotheses, we also get the asymptotic normality of the sequence of estimators $\widehat{\theta}_n$ of $\theta_0$. We denote by $\nabla_{00} Q_n(\theta)$ the Hessian matrix of the functional $Q_n$ evaluated at $\theta$.

\begin{The} (Theorem 3.1 in \cite{MR1315971}) \\
	Suppose that $\widehat{\theta}_n \to \theta_0$ in probability,  (i) $\theta_0$ is an interior point of $\Theta$, (ii) $Q_n(\theta)$ is twice differentiable in a neighborhood $\mathcal{N}$ of $\theta_0$, (iii) $\sqrt{n} \nabla_0 Q_n(\theta_0) \rightsquigarrow \mathcal{N}\left( 0, \boldsymbol{\Sigma} \right)$, (iv) there is $\mathbf{H}(\theta)$ continuous at $\theta_0$ and $\sup_{\theta \in \mathcal{N}} \| \nabla_{00} Q_n(\theta) - \mathbf{H}(\theta)  \| \to 0$ in probability (v) $\mathbf{H} = \mathbf{H}(\theta_0)$ is non singular. 
	Then 
	\[
	\sqrt{n} \left( \widehat{\theta}_n - \theta_0 \right) \rightsquigarrow \mathcal{N}\left( 0, \mathbf{H}^{-1}\boldsymbol{\Sigma}\mathbf{H}^{-1}\right)
	\]   
	\label{the:newey_macfadden_3}
\end{The}

We also recall the central limit theorem for bootstrap means (see Theorem 23.4 in \cite{MR1652247} for a proof). 
\begin{The} (Central limit theorem for the bootstrap means) \\
	Let $X_1, X_2, \ldots$ be i.i.d. random vectors with mean $\boldsymbol{\mu}$ and covariance matrix $\boldsymbol{\Gamma}$. Then conditionally on $X_1, X_2, \ldots$, for almost every sequence $X_1, X_2, \ldots$
	\[
	\sqrt{n} \left( \overline{X}^*_n -  \overline{X}_n \right) \rightsquigarrow \mathcal{N}\left( 0, \boldsymbol{\Gamma} \right)
	\]   
	\label{the:bmclt}
	where $\overline{X}_n$ is the empirical mean and $\overline{X}^*_n$ is the empirical mean of $n$ independent observations drawn from the empirical distribution.
\end{The}	

\section{Proofs}\label{sec:proofs}

\begin{proof}{\em of Lemma \ref{lem:mm_consist}} \\
	First note that the assumptions $\E(Y^2) < + \infty$ and $\E(\| \mathbf{X}\| ^2) < + \infty$  ensure the existence of $\sigma_Y^2$ and  $\boldsymbol{\Gamma}_X$.  
	From the  law of large numbers, we have  that for all $k \in \{1, \ldots, K\}, \ \widehat{\boldsymbol{\mu}}_{1,X}^k \to \boldsymbol{\mu}_{X}^k$ and $\widehat{\mu}_{Y}^k \to \mu_{Y}^k$ in probability when $n_{\min}$ tends to  infinity. 
	
	We deduce from  the continuous mapping theorem that $\widehat{\boldsymbol{\mu}}_{1,X}^\top  \widehat{\boldsymbol{\mu}}_{1,X} \to   \boldsymbol{\mu}_{1,X}^\top \mathbf{w} \boldsymbol{\mu}_{1,X}$  and  $\widehat{\boldsymbol{\mu}}_{1,X}^\top \mathbf{w} \widehat{\boldsymbol{\mu}}_Y \to   \boldsymbol{\mu}_{1,X}^\top \mathbf{w} \boldsymbol{\mu}_Y$ in probability. 
	Under hypothesis $\mathbf{H}_1$, the inverse being continuous in a neighborhood of $\boldsymbol{\mu}_{1,X}^\top \mathbf{w} \boldsymbol{\mu}_{1,X}$ another application of the continuous mapping theorem gives that $\left( \widehat{\boldsymbol{\mu}}_{1,X}^\top \mathbf{w} \widehat{\boldsymbol{\mu}}_{1,X}\right)^{-1} \to \left( \boldsymbol{\mu}_{1,X}^\top \mathbf{w} \boldsymbol{\mu}_{1,X} \right)^{-1}$ and 
	\[
	\widehat{\boldsymbol{\beta}}^M =  \left( \widehat{\boldsymbol{\mu}}_{1,X}^\top \mathbf{w} \widehat{\boldsymbol{\mu}}_{1,X}\right)^{-1} \widehat{\boldsymbol{\mu}}_{1,X}^\top \mathbf{w} \widehat{\boldsymbol{\mu}}_Y  \to  \left( \boldsymbol{\mu}_{1,X}^\top \mathbf{w} \boldsymbol{\mu}_{1,X} \right)^{-1} \boldsymbol{\mu}_{1,X}^\top \mathbf{w} \boldsymbol{\mu}_Y = \boldsymbol{\beta}
	\]
	in probability as $n_{\min}$ tends to infinity.

	The law of large numbers gives that $\widehat{\boldsymbol{\Gamma}}_X^2 \to \boldsymbol{\Gamma}_X^2$ and $\widehat{\sigma}_Y^2 \to \sigma_Y^2$ in probability and we deduce, with another application of the continuous mapping theorem,  that $\widehat{\sigma}_Y^2 - \widehat{\boldsymbol{\beta}}^\top \widehat{\boldsymbol{\Gamma}}_X^2 \widehat{\boldsymbol{\beta}} \to \sigma_Y^2 - \boldsymbol{\beta}^\top \boldsymbol{\Gamma}_X^2 \boldsymbol{\beta}
	= \sigma_\epsilon^2$ in probability as $n_{\min}$ tends to infinity.
\end{proof}

\begin{proof}{\em of Lemma \ref{lem:ot_consist}} \\
	The proof is based on Lemma 2.9  and Theorem 2.1 in \cite{MR1315971}, which are recalled in Appendix \ref{sec:theorems}. 
	
	The law of large numbers and the continuous mapping theorem give us that for all $(\boldsymbol{\gamma}, \sigma^2_{\gamma}) \in \Theta$, $\varphi_n(\boldsymbol{\gamma}, \sigma^2_{\gamma}) \to \varphi(\boldsymbol{\gamma}, \sigma^2_{\gamma})$ in probability, when $n_{\min}$ tends to infinity.
	
	\begin{comment}			
		\begin{align}
			\varphi_n(\boldsymbol{\gamma},\sigma^2) &= \sum_{k=1}^K \pi_k \left[ 
			(\widehat{\mu}_{Y}^k - \gamma_0 - \boldsymbol{\gamma}_{-0}^\top \widehat{\boldsymbol{\mu}}_{X}^k)^2 +  \left(\widehat{\sigma}_{Y,k} - \sqrt{\boldsymbol{\gamma}_{-0}^\top \widehat{\boldsymbol{\Gamma}}_X^k  \boldsymbol{\gamma}_{-0}+ \sigma^2} \right)^2
			\right].
			\label{eq:wass_gn}
		\end{align}
	\end{comment}
	
	Consider now $(\boldsymbol{\alpha}, \sigma_\alpha^2) \in \Theta$. We have, 
	\begin{align*}
		\hspace{-15pt}
		\left| (\widehat{\mu}_{Y}^k - \gamma_0 - \boldsymbol{\gamma}_{-0}^\top \widehat{\boldsymbol{\mu}}_{X}^k)^2 - 
		(\widehat{\mu}_{Y}^k - \alpha_0 - \boldsymbol{\alpha}_{-0}^\top \widehat{\boldsymbol{\mu}}_{X}^k)^2 \right| & = 
		\left| (\boldsymbol{\alpha} - \boldsymbol{\gamma})^T  \begin{pmatrix} 1 \\ \widehat{\boldsymbol{\mu}}_{X}^k \end{pmatrix} \left(2 \widehat{\mu}_{Y}^k - (\boldsymbol{\alpha} + \boldsymbol{\gamma})^T \begin{pmatrix} 1 \\ \widehat{ \boldsymbol{\mu}}_{X}^k  \end{pmatrix}\right) \right| \\
		&\leq  \norm{\boldsymbol{\alpha} - \boldsymbol{\gamma}} A_{n}^k,
	\end{align*}
	with Cauchy-Schwarz inequality and $A_{n,k} = \mathcal{O}_p(1)$ because  $\| \widehat{\boldsymbol{\mu}}_{X}^k \| = O_p(1)$, $\widehat{\mu}_{Y}^k= O_p(1)$ and for some constant $C_{1}$ that does not depend on $\boldsymbol{\alpha}$ and $\boldsymbol{\gamma}$, $\|\boldsymbol{\alpha} + \boldsymbol{\gamma}\| \leq C_{1} < \infty$ because $\Theta$ is supposed to be compact. 
	
	On the other hand, we have
	\begin{align*}
		& \left| \left(\widehat{\sigma}_{Y,k} - \sqrt{\boldsymbol{\gamma}_{-0}^\top \widehat{\boldsymbol{\Gamma}}_X^k  \boldsymbol{\gamma}_{-0}+ \sigma_\gamma^2} \right)^2 - \left(\widehat{\sigma}_{Y,k} - \sqrt{\boldsymbol{\alpha}_{-0}^\top \widehat{\boldsymbol{\Gamma}}_X^k  \boldsymbol{\alpha}_{-0}+ \sigma_\alpha^2} \right)^2 \right|  \\ 
		& = \left| \sqrt{\boldsymbol{\alpha}_{-0}^\top \widehat{\boldsymbol{\Gamma}}_X^k  \boldsymbol{\alpha}_{-0}+ \sigma_\alpha^2} -\sqrt{\boldsymbol{\gamma}_{-0}^\top \widehat{\boldsymbol{\Gamma}}_X^k  \boldsymbol{\gamma}_{-0}+ \sigma_\gamma^2} \right|
		\left(  2\widehat{\sigma}_{Y,k} +\sqrt{\boldsymbol{\alpha}_{-0}^\top \widehat{\boldsymbol{\Gamma}}_X^k  \boldsymbol{\alpha}_{-0}+ \sigma_\alpha^2}  + \sqrt{\boldsymbol{\gamma}_{-0}^\top \widehat{\boldsymbol{\Gamma}}_X^k  \boldsymbol{\gamma}_{-0}+ \sigma_\gamma^2} \right) \\
		&= \left| \sqrt{\boldsymbol{\alpha}_{-0}^\top \widehat{\boldsymbol{\Gamma}}_X^k  \boldsymbol{\alpha}_{-0}+ \sigma_\alpha^2} -\sqrt{\boldsymbol{\gamma}_{-0}^\top \widehat{\boldsymbol{\Gamma}}_X^k  \boldsymbol{\gamma}_{-0}+ \sigma_\gamma^2} \right| O_p(1) 
		%\\
		%\left|  \sqrt{\beta_1^2  \widehat{ \sigma_{X,k}^2} + \sigma_\epsilon^2} -  \sqrt{\alpha_1^2  \widehat{ \sigma_{X,k}^2} + \sigma_\epsilon^2} \right| 
		%&=\left|  \sqrt{  \beta^T \begin{pmatrix} 0 \\ \widehat{ \sigma_{X,k}^2}  \\1 \end{pmatrix}} -  \sqrt{  \alpha^T \begin{pmatrix} 0 \\ \widehat{ \sigma_{X,k}^2}  \\1 \end{pmatrix}} \right|  \\
		%&\leq  \cfrac{ \left[ \beta - \alpha \right]}{2\sqrt{ \alpha^T \begin{pmatrix} 0 \\ \widehat{ \sigma_{X,k}^2}  \\1 \end{pmatrix}}} \leq   \norm{ \alpha - \beta} \mathcal{O}_p(1),
	\end{align*}
	since $\Theta$ is compact and  $\|\widehat{\boldsymbol{\Gamma}}_X^k \|_{sp} = O_p(1)$, where $\|.\|_{sp}$ denotes the spectral norm.  Because  $\boldsymbol{\alpha}_{-0}^\top \widehat{\boldsymbol{\Gamma}}_X^k  \boldsymbol{\alpha}_{-0} - \boldsymbol{\gamma}_{-0}^\top \widehat{\boldsymbol{\Gamma}}_X^k  \boldsymbol{\gamma}_{-0} = \boldsymbol{\alpha}_{-0}^\top \widehat{\boldsymbol{\Gamma}}_X^k  \left( \boldsymbol{\alpha}_{-0} -  \boldsymbol{\gamma}_{-0} \right) + \left( \boldsymbol{\alpha}_{-0} -  \boldsymbol{\gamma}_{-0} \right)^\top \widehat{\boldsymbol{\Gamma}}_X^k  \boldsymbol{\gamma}_{-0}$ we have, for some constant $C_{2,k} >0$,
	\begin{align}
		\left| \boldsymbol{\alpha}_{-0}^\top \widehat{\boldsymbol{\Gamma}}_X^k  \boldsymbol{\alpha}_{-0} - \boldsymbol{\gamma}_{-0}^\top \widehat{\boldsymbol{\Gamma}}_X^k  \boldsymbol{\gamma}_{-0}   \right|  & \leq C_{2,k} \left\| \widehat{\boldsymbol{\Gamma}}_X^k \right\|_{sp} \| \boldsymbol{\alpha} - \boldsymbol{\gamma} \|.  
	\end{align}
	Using now the fact that function $x \mapsto \sqrt{x}$ is concave and differentiable, we have  for $x>0$ and $y>0$ that $\sqrt{y} \leq \sqrt{x} + \frac{y-x}{2\sqrt{x}}$. Thus, if $y>x>0$ then  $0 < \sqrt{y} - \sqrt{x} \leq  \frac{y-x}{2\sqrt{x}}$ and if $x>y>0$, then $0 < \sqrt{x} - \sqrt{y} \leq  \frac{x-y}{2\sqrt{y}}$. Consequently, we have $|\sqrt{y} - \sqrt{x} | \leq \frac{|x-y|}{2 \min(\sqrt{x},\sqrt{y})}$ and we deduce that,
	\begin{align}
		\left| \sqrt{\boldsymbol{\alpha}_{-0}^\top \widehat{\boldsymbol{\Gamma}}_X^k  \boldsymbol{\alpha}_{-0}+ \sigma_\alpha^2} -\sqrt{\boldsymbol{\gamma}_{-0}^\top \widehat{\boldsymbol{\Gamma}}_X^k  \boldsymbol{\gamma}_{-0}+ \sigma_\gamma^2} \right|  & \leq   B_n^k \left(  \| \boldsymbol{\alpha} - \boldsymbol{\gamma} \|  + | \sigma_\alpha^2 - \sigma_\gamma^2 | \right)
	\end{align}
	where $B_n^k = O_p(1)$.
	
	Combining previous inequalities, we get
	\begin{align}
		\left|  \varphi_n(\boldsymbol{\gamma}, \sigma^2_{\gamma}) - \varphi_n(\boldsymbol{\alpha}, \sigma_\alpha^2) \right| & \leq  \left(  \| \boldsymbol{\alpha} - \boldsymbol{\gamma} \|  + | \sigma_\alpha^2 - \sigma_\gamma^2 | \right)
		\sum_{k=1}^K \pi_k \left( B_n^k + A_n^k \right),
		\label{ineq:deltaphin}	
	\end{align} 
	with $\sum_{k=1}^K \pi_k \left( B_n^k + A_n^k \right) = O_p(1)$. 
	As a result, it can be deduced from Lemma 2.9 in  \cite{MR1315971} that
	\[
	\sup_{ (\boldsymbol{\gamma}, \sigma^2_\gamma) \in \Theta}  \left| \varphi_n(\boldsymbol{\gamma}, \sigma^2_{\gamma}) - \varphi(\boldsymbol{\gamma}, \sigma^2_{\gamma})\right| \to 0  \quad \mbox{ in probability.}
	\] 
	We conclude the proof by recalling that $\varphi(\boldsymbol{\gamma}, \sigma^2_{\gamma})$ attains its unique minimum at $(\boldsymbol{\beta}, \sigma^2_{\epsilon}) \in \Theta$ if assumption $\mathbf{H}_1$ is fulfilled, so that  $(\widehat{\boldsymbol{\beta}}^W, \widehat{\sigma}^{2,W}) \to (\boldsymbol{\beta}, \sigma^2_{\epsilon})$ in probability in view of Theorem 2.1 in \cite{MR1315971}.
\end{proof}

\begin{proof}{\em of Proposition \ref{prop:mm_asympt_norm}}

	The central limit theorem applies directly to the independent sequences of independent random variables $(\mathbf{X}_1^1, \cdots, \mathbf{X}_n^1), \ldots, (\mathbf{X}_1^K, \cdots, \mathbf{X}_n^K)$ and $(Y_1^1, \cdots, Y_n^1), , \ldots, (Y_1^K, \cdots, Y_n^K)$ so that, as $n$ tends to infinity
	\begin{align}
		\sqrt{n} \begin{pmatrix} \widehat{\boldsymbol{\mu}}_{X}^1  - \boldsymbol{\mu}_{X}^1 \\
			\vdots \\ \widehat{\boldsymbol{\mu}}_{X}^K  - \boldsymbol{\mu}_{X}^K \\
			\widehat{\mu}_{Y}^1  - \mu_{Y}^1 \\
			\vdots \\ \widehat{\mu}_{Y}^K  - \mu_{Y}^K 
		\end{pmatrix}
		&\rightsquigarrow \mathcal{N}\left( 0, \boldsymbol{\Gamma}_{\mu} \right)
		\label{prop:asn1}
	\end{align}
	where $ \boldsymbol{\Gamma}_{\mu}$ is a block diagonal matrix, with diagonal elements $(\boldsymbol{\Gamma}_{X}^1, \ldots, \boldsymbol{\Gamma}_{X}^K, \sigma_{Y,1}^2, \ldots, \sigma_{Y,K}^2)$, with $\boldsymbol{\Gamma}_{X}^k = \mbox{Var}( \mathbf{X} | G=k) = \E \left( \mathbf{X}^k (\mathbf{X}^k)^\top \right) - \boldsymbol{\mu}_{X}^k (\boldsymbol{\mu}_{X}^k)^\top$ and $\sigma_{Y,k}^2 = \mbox{Var}( Y | G=k)$. % \boldsymbol{\beta}^\top \boldsymbol{\Gamma}_{X}^k \boldsymbol{\beta}^\top + \sigma_{\epsilon}^2$.
	Consider the application $g : \mathbb{R}^{dK+K} \to \mathbb{R}^{d+1}$ defined by 
	\[
	g( \boldsymbol{\mu}_{X}^1, \ldots,  \boldsymbol{\mu}_{X}^K, \mu_{Y}^1, \ldots, \mu_{Y}^K) = \left( \boldsymbol{\mu}_{1,X}^\top \mathbf{w} \boldsymbol{\mu}_{1,X} \right)^{-1} \boldsymbol{\mu}_{1,X}^\top \mathbf{w} \boldsymbol{\mu}_Y.
	\]
	Application $g$ is differentiable at $\boldsymbol{\theta}  = (\boldsymbol{\mu}_{X}^1, \ldots,  \boldsymbol{\mu}_{X}^K, \mu_{Y}^1, \ldots, \mu_{Y}^K)$,  with non null Jacobian matrix denoted by $\mathbf{J}_\theta$ (see Chapter 8 and more particularly Theorem 8.3 in \cite{Magnus_Neudecker}).  The  application of the Delta method (see Theorem 3.1 in \cite{MR1652247}) allows to get the asymptotic normality convergence result,
	\begin{align*}
		\sqrt{n} \left( \widehat{\boldsymbol{\beta}}^M - \boldsymbol{\beta} \right) 
		&\rightsquigarrow \mathcal{N}\left(0, \boldsymbol{\Gamma}_{\beta_M} \right),
	\end{align*}
	where $\boldsymbol{\Gamma}_{\beta_M} = \mathbf{J}_\theta \boldsymbol{\Gamma}_{\mu} \mathbf{J}_\theta^\top$.
\end{proof}

\begin{proof}{\em of Proposition \ref{prop:ot_asympt_norm}}
	
	The proof consists in checking the different points of Theorem \ref{the:newey_macfadden_3}. Point (i) is satisfied by the assumptions, and the point (ii) follows directly from the fact that $\varphi_n(\boldsymbol{\gamma}, \sigma^2)$ is twice-differentiable in a neighborhood of  $(\boldsymbol{\beta}, \sigma_\epsilon^2)$. To show that (iii) is fulfilled, we consider the following expansion, based on the empirical version of the gradient of $\varphi$:
	\begin{align}
		\nabla \varphi_n & = \begin{pmatrix} - 2 \sum_{k=1}^K \pi_k \left( \widehat{\mu}_{Y}^k - \beta_0 - \boldsymbol{\beta}_{-0}^\top \widehat{\boldsymbol{\mu}}_{X}^k\right) \\ - 2 \sum_{k=1}^K \pi_k \left[ \left( \widehat{\mu}_{Y}^k - \beta_0 - \boldsymbol{\beta}_{-0}^\top \widehat{\boldsymbol{\mu}}_{X}^k\right)\widehat{\boldsymbol{\mu}}_{X}^k + \left(\frac{ \widehat{\sigma}_{Y,k}}{\sqrt{ \boldsymbol{\beta}_{-0}^\top \widehat{\boldsymbol{\Gamma}}_X^k\boldsymbol{\beta}_{-0} + \sigma_\epsilon^2}}  - 1 \right)  \widehat{\boldsymbol{\Gamma}}_X^k  \boldsymbol{\beta}_{-0} 
			\right]
			\\ \sum_{k=1}^K \pi_k \left(1-\frac{\widehat{\sigma}_{Y,k}}{\sqrt{ \boldsymbol{\beta}_{-0}^\top \widehat{\boldsymbol{\Gamma}}_X^k  \boldsymbol{\beta}_{-0} + \sigma_\epsilon^2}} \right)
		\end{pmatrix}
		\label{def:gradvarphi_n}
	\end{align}
	Since model (\ref{def:lmm}) holds, $\nabla \varphi =0$ and  $\widehat{\mu}_{Y}^k - \beta_0 - \boldsymbol{\beta}_{-0}^\top \widehat{\boldsymbol{\mu}}_{X}^k = (\widehat{\mu}_{Y}^k - {\mu}_{Y}^k) - \boldsymbol{\beta}_{-0}^\top \left(\widehat{\boldsymbol{\mu}}_{X}^k - \boldsymbol{\mu}_{X}^k \right)$, we thus deduce with (\ref{prop:asn1}) the asymptotic normality of the first component of the gradient $\nabla \varphi_n$, that is to say $\sqrt{n} \left(- 2 \sum_{k=1}^K \pi_k \left( \widehat{\mu}_{Y}^k - \beta_0 - \boldsymbol{\beta}_{-0}^\top \widehat{\boldsymbol{\mu}}_{X}^k\right) \right)$ converges in distribution to a centered Gaussian distribution. As far as the second component is concerned, it can be noted that $\widehat{\boldsymbol{\Gamma}}_X^k$ converges in probability to $\boldsymbol{\Gamma}_X^k$, and by the continuous mapping theorem $\sqrt{ \boldsymbol{\beta}_{-0}^\top \widehat{\boldsymbol{\Gamma}}_X^k  \boldsymbol{\beta}_{-0} + \sigma_\epsilon^2} \to \sigma_{Y,k}$ in probability. It can also be noted that, under the moment condition $\E \left[ \| \mathbf{X} \|^4 | G = k \right] < \infty$, the central limit theorem gives that $\sqrt{n}\left( \widehat{\boldsymbol{\Gamma}}_X^k - \boldsymbol{\Gamma}_X^k \right)$ converges in distribution to a centered Gaussian multivariate distribution, and we deduce with the Cramer-Wold device, the continuous mapping theorem and Slutsky's theorem that the second component of $\nabla \varphi_n$ multiplied by $\sqrt{n}$ also in distribution to a centered Gaussian random vector. It is immediate to deduce that the same convergence result holds for the third component, which is to say that $\sqrt{n}\left( \sum_{k=1}^K \pi_k \left(1-\frac{\widehat{\sigma}_{Y,k}}{\sqrt{ \boldsymbol{\beta}_{-0}^\top \widehat{\boldsymbol{\Gamma}}_X^k  \boldsymbol{\beta}_{-0} + \sigma_\epsilon^2}}\right)\right)$ converges in distribution to a centered Gaussian random variable.
	We finally deduce, with the Cramer-Wold device, that  (iii) is fulfilled. 
	
	\smallskip
	
	To prove that (iv) also holds, consider the Hessian matrix of functional $\varphi_n$, evaluated at $(\boldsymbol{\beta}, \sigma_\epsilon^2)$:
	\begin{footnotesize}
		\begin{align*}
			\hspace{-15pt}
			\nabla_{00}\varphi_n & = \begin{pmatrix}
				2 & 2 \left(\sum_{k=1}^K \pi_k \widehat{\boldsymbol{\mu}}_{X}^k \right)^\top & 0 \\
				2 \sum_{k=1}^K \pi_k \widehat{\boldsymbol{\mu}}_{X}^k & \widehat{\mathbf{H}}(\boldsymbol{\beta}_{-0}) & \sum_{k=1}^K \pi_k \widehat{\sigma}_{Y,k} \left( \boldsymbol{\beta}_{-0}^\top \widehat{\boldsymbol{\Gamma}}_X^k  \boldsymbol{\beta}_{-0} + \sigma_\epsilon^2 \right)^{-3/2} \widehat{\boldsymbol{\Gamma}}_X^k \boldsymbol{\beta}_{-0} \\
				0 & \sum_{k=1}^K \pi_k \widehat{\sigma}_{Y,k} \left( \boldsymbol{\beta}_{-0}^\top \widehat{\boldsymbol{\Gamma}}_X^k  \boldsymbol{\beta}_{-0} + \sigma_\epsilon^2 \right)^{-3/2} \left(\widehat{\boldsymbol{\Gamma}}_X^k \boldsymbol{\beta}_{-0}\right)^\top
				& \frac{1}{2} \sum_{k=1}^K \pi_k \widehat{\sigma}_{Y,k} \left( \boldsymbol{\beta}_{-0}^\top \widehat{\boldsymbol{\Gamma}}_X^k  \boldsymbol{\beta}_{-0} + \sigma_\epsilon^2 \right)^{-3/2} \end{pmatrix},
		\end{align*}
	\end{footnotesize}
	where
	\begin{align*}
		\hspace{-18pt}
		\widehat{\mathbf{H}}(\boldsymbol{\beta}_{-0}) 
		=& 2 \sum_{k=1}^K \pi_k \bigg[
		\widehat{\sigma}_{Y,k} \left( \boldsymbol{\beta}_{-0}^\top \widehat{\boldsymbol{\Gamma}}_X^k  \boldsymbol{\beta}_{-0} + \sigma_\epsilon^2 \right)^{-3/2} 
		\bigg[
		\left( \widehat{\boldsymbol{\Gamma}}_X^k  \boldsymbol{\beta}_{-0} \right) \left( \widehat{\boldsymbol{\Gamma}}_X^k  \boldsymbol{\beta}_{-0} \right)^\top
		- \left( \boldsymbol{\beta}_{-0}^\top \widehat{\boldsymbol{\Gamma}}_X^k  \boldsymbol{\beta}_{-0} + \sigma_\epsilon^2 \right) \widehat{\boldsymbol{\Gamma}}_X^k
		\bigg] \\ 
		&+\widehat{\boldsymbol{\mu}}_{X}^k \left(\widehat{\boldsymbol{\mu}}_{X}^k \right)^\top + \widehat{\boldsymbol{\Gamma}}_X^k 
		\bigg].
	\end{align*}
	By similar arguments as those used to show that $\varphi_n (\boldsymbol{\beta}, \sigma_\epsilon^2)$ converges in probability to  
	$\varphi(\boldsymbol{\beta}, \sigma_\epsilon^2)$, we deduce that $\nabla_{00}\varphi_n$ converges in probability to some matrix $\mathbf{H}(\boldsymbol{\beta}, \sigma_\epsilon^2)$, defined as follows
	\begin{footnotesize}
		\begin{align*}
			\hspace{-15pt}
			\mathbf{H}(\boldsymbol{\beta}, \sigma_\epsilon^2) & = \begin{pmatrix}
				2 & 2 \left(\sum_{k=1}^K \pi_k \boldsymbol{\mu}_{X}^k \right)^\top & 0 \\
				2 \sum_{k=1}^K \pi_k \boldsymbol{\mu}_{X}^k & \mathbf{H}(\boldsymbol{\beta}_{-0}) & \sum_{k=1}^K \pi_k \sigma_{Y,k} \left( \boldsymbol{\beta}_{-0}^\top \boldsymbol{\Gamma}_X^k  \boldsymbol{\beta}_{-0} + \sigma_\epsilon^2 \right)^{-3/2}  \boldsymbol{\Gamma}_X^k \boldsymbol{\beta}_{-0} \\
				0 & \sum_{k=1}^K \pi_k \sigma_{Y,k} \left( \boldsymbol{\beta}_{-0}^\top \boldsymbol{\Gamma}_X^k  \boldsymbol{\beta}_{-0} + \sigma_\epsilon^2 \right)^{-3/2} \left(\boldsymbol{\Gamma}_X^k \boldsymbol{\beta}_{-0}\right)^\top
				& \frac{1}{2} \sum_{k=1}^K \pi_k \sigma_{Y,k} \left( \boldsymbol{\beta}_{-0}^\top \boldsymbol{\Gamma}_X^k  \boldsymbol{\beta}_{-0} + \sigma_\epsilon^2 \right)^{-3/2} \end{pmatrix}
		\end{align*}
	\end{footnotesize}
	where 
	\begin{align*}
		\hspace{-13pt}
		\mathbf{H}(\boldsymbol{\beta}_{-0}) 
		=& 2 \sum_{k=1}^K \pi_k \bigg(
		\sigma_{Y,k} \left( \boldsymbol{\beta}_{-0}^\top \boldsymbol{\Gamma}_X^k  \boldsymbol{\beta}_{-0} + \sigma_\epsilon^2 \right)^{-3/2} 
		\bigg[
		\left( \boldsymbol{\Gamma}_X^k  \boldsymbol{\beta}_{-0} \right) \left( \boldsymbol{\Gamma}_X^k  \boldsymbol{\beta}_{-0} \right)^\top
		- \left( \boldsymbol{\beta}_{-0}^\top \boldsymbol{\Gamma}_X^k  \boldsymbol{\beta}_{-0} + \sigma_\epsilon^2 \right) \boldsymbol{\Gamma}_X^k
		\bigg] \\ 
		&+ \boldsymbol{\mu}_{X}^k \left(\boldsymbol{\mu}_{X}^k \right)^\top + \boldsymbol{\Gamma}_X^k 
		\bigg).
	\end{align*}
	
	We now must check that $\mathbf{H}(\boldsymbol{\beta}, \sigma_\epsilon^2)$ is a positive definite matrix. For that we show that at the minimizer value $(\boldsymbol{\beta}, \sigma_\epsilon^2)$ its determinant is strictly positive.
	We first note that $\sigma_{Y,k} = \left( \boldsymbol{\beta}_{-0}^\top \boldsymbol{\Gamma}_X^k  \boldsymbol{\beta}_{-0} + \sigma_\epsilon^2\right) ^{1/2}$ so that 	$ \sigma_{Y,k} \left( \boldsymbol{\beta}_{-0}^\top \boldsymbol{\Gamma}_X^k  \boldsymbol{\beta}_{-0} + \sigma_\epsilon^2 \right)^{-3/2} =\frac{1}{\sigma_{Y,k}^2}$ and $\mathbf{H}(\boldsymbol{\beta}_{-0})$ can be written in a simpler form, 
	\begin{align}
		\mathbf{H}(\boldsymbol{\beta}_{-0}) 
		=& 2 \sum_{k=1}^K \pi_k \left[ \boldsymbol{\mu}_{X}^k \left(\boldsymbol{\mu}_{X}^k \right)^\top + \frac{1}{\sigma_{Y,k}^2} \boldsymbol{\Gamma}_X^k  \boldsymbol{\beta}_{-0} \left(\boldsymbol{\Gamma}_X^k  \boldsymbol{\beta}_{-0}\right)^\top  \right],
	\end{align}
	which is a positive definite matrix under the hypothesis $\mathbf{H}_1$. Using a block matrix determinant formula, we have
	\begin{align}
		\left|  \mathbf{H}(\boldsymbol{\beta}, \sigma_\epsilon^2)\right| & = \begin{vmatrix} 2 & 0 \\ 0 & \frac{1}{2} \sum_{k=1}^K \frac{\pi_k}{\sigma_{Y,k}^2} \end{vmatrix} \ \begin{vmatrix} \mathbf{H}(\boldsymbol{\beta}_{-0})  -  \mathbf{C}  \begin{pmatrix} \frac{1}{2} & 0 \\ 0 & \frac{2}{\sum_k \frac{\pi_k}{\sigma_{Y,k}^2}} \end{pmatrix}  \mathbf{C}^\top \end{vmatrix} 
		\label{det:H}
	\end{align} 					
	where $\mathbf{C} = \begin{pmatrix} 2 \sum_{k=1}^K \pi_k \boldsymbol{\mu}_{X}^k &  \sum_{k=1}^K \frac{\pi_k}{ \sigma_{Y,k}^2}  \boldsymbol{\Gamma}_X^k \boldsymbol{\beta}_{-0} \end{pmatrix}$, and it only has to be verified that the second determinant at the righthand side of (\ref{det:H}) is strictly positive. We now have to show that
	\begin{align}
		\mathbf{H}(\boldsymbol{\beta}_{-0})  -  \mathbf{C}  \begin{pmatrix} \frac{1}{2} & 0 \\ 0 & \frac{2}{\sum_k \frac{\pi_k}{\sigma_{Y,k}^2}} \end{pmatrix}  \mathbf{C}^\top  = 2 \sum_{k=1}^K \pi_k \boldsymbol{\mu}_{X}^k \left(\boldsymbol{\mu}_{X}^k \right)^\top  - 2  \left(\sum_{k=1}^K \pi_k \boldsymbol{\mu}_{X}^k \right)\left(\sum_{k=1}^K \pi_k \boldsymbol{\mu}_{X}^k \right)^\top  \nonumber \\
		+  2 \sum_{k=1}^K \frac{\pi_k}{\sigma_{Y,k}^2} \boldsymbol{\Gamma}_X^k  \boldsymbol{\beta}_{-0} \left(\boldsymbol{\Gamma}_X^k  \boldsymbol{\beta}_{-0}\right)^\top - \frac{2}{\sum_k \frac{\pi_k}{\sigma_{Y,k}^2}} \left(\sum_{k=1}^K \frac{\pi_k}{ \sigma_{Y,k}^2}  \boldsymbol{\Gamma}_X^k \boldsymbol{\beta}_{-0}\right)\left(\sum_{k=1}^K \frac{\pi_k}{ \sigma_{Y,k}^2}  \boldsymbol{\Gamma}_X^k \boldsymbol{\beta}_{-0}\right)^\top
		\label{Hdefipo}
	\end{align} 
	is a positive matrix.
	We can remark that by Cauchy Schwarz inequality, for $\mathbf{u}  \in \mathbb{R}^d$, 
	\begin{align*}
		\mathbf{u}^\top  \left(\sum_{k=1}^K \pi_k \boldsymbol{\mu}_{X}^k \right)\left(\sum_{k=1}^K \pi_k \boldsymbol{\mu}_{X}^k \right)^\top \mathbf{u} =   \left(\sum_{k=1}^K \pi_k \mathbf{u}^\top\boldsymbol{\mu}_{X}^k \right)^2  
		& \leq  \sum_{k=1}^K \pi_k  \left( \mathbf{u}^\top \boldsymbol{\mu}_{X}^k\right)^2 \\
		& = \mathbf{u}^\top \left( \sum_{k=1}^K \pi_k  \boldsymbol{\mu}_{X}^k \left(\boldsymbol{\mu}_{X}^k \right)^\top \right)\mathbf{u} 
	\end{align*}
	using the fact that $\sum_k (\sqrt{\pi_k})^2 = 1$. It can be noted that if $\mathbf{u} \neq 0$, previous inequality is strict unless $ \mathbf{u}^\top \boldsymbol{\mu}_{X}^1 = \cdots =\mathbf{u}^\top \boldsymbol{\mu}_{X}^K$, which cannot happen under the hypothesis $\mathbf{H}_1$. The second part at the righthand side of (\ref{Hdefipo}) is handled in the same way. We have
	\begin{align*}
		\mathbf{u}^\top \left(\sum_{k=1}^K \frac{\pi_k}{ \sigma_{Y,k}^2}  \boldsymbol{\Gamma}_X^k \boldsymbol{\beta}_{-0}\right)\left(\sum_{k=1}^K \frac{\pi_k}{ \sigma_{Y,k}^2}  \boldsymbol{\Gamma}_X^k \boldsymbol{\beta}_{-0}\right)^\top\mathbf{u} & = \left( \mathbf{u}^\top \left(\sum_{k=1}^K \frac{\pi_k}{ \sigma_{Y,k}^2}  \boldsymbol{\Gamma}_X^k \boldsymbol{\beta}_{-0}\right) \right)^2 \\
		& \leq \sum_{k=1}^K \sqrt{\frac{\pi_k}{ \sigma_{Y,k}^2}}^2  \sum_{k=1}^K \left(\sqrt{\frac{\pi_k}{ \sigma_{Y,k}^2}}^2  \mathbf{u}^\top\boldsymbol{\Gamma}_X^k \boldsymbol{\beta}_{-0} \right)^2 \\
		&=  \sum_{k=1}^K \frac{\pi_k}{ \sigma_{Y,k}^2} \sum_{k=1}^K \frac{\pi_k}{ \sigma_{Y,k}^2} \mathbf{u}^\top \boldsymbol{\Gamma}_X^k  \boldsymbol{\beta}_{-0} \left(\boldsymbol{\Gamma}_X^k  \boldsymbol{\beta}_{-0}\right)^\top \mathbf{u},
	\end{align*}
	and consequently the determinant of $\mathbf{H}(\boldsymbol{\beta}, \sigma_\epsilon^2)$ is strictly positive.
	
	To finish the proof, it remains to check that  in a neighborhood $\mathcal{N}$ of $(\boldsymbol{\beta}, \sigma_\epsilon^2)$,  we have 
	\[
	\sup_{(\boldsymbol{\gamma}, \sigma^2_{\gamma}) \in \mathcal{N}} \| \nabla_{00} \varphi_n(\boldsymbol{\gamma}, \sigma^2_{\gamma}) - \mathbf{H}(\boldsymbol{\gamma}, \sigma^2_{\gamma})  \| \to 0 \ \mbox{ in probability.} 
	\]
	This is a direct consequence of the continuous mapping theorem, which gives us that for all  $(\boldsymbol{\gamma}, \sigma^2_{\gamma}) \in \mathcal{N}$, $\| \nabla_{00} \varphi_n(\boldsymbol{\gamma}, \sigma^2_{\gamma}) - \mathbf{H}(\boldsymbol{\gamma}, \sigma^2_{\gamma})  \| \to 0$ in probability, and the fact that third order partial derivatives of $\varphi_n(\boldsymbol{\gamma}, \sigma^2_{\gamma})$ are bounded in probability for $(\boldsymbol{\gamma}, \sigma^2_{\gamma})$ so that Theorem~\ref{the:newey_macfadden_1} can apply.
\end{proof}

\begin{proof}{\em of Lemma \ref{lem:d=1}} \\
	Note that  
	\[
	\widehat{\beta}_1 = g(\widehat{\mu}_{X}^1, \ldots, \widehat{\mu}_{X}^K, \widehat{\mu}_{Y}^1, \ldots, \widehat{\mu}_{Y}^K),
	\]
	with  $g : \mathbb{R}^{K+K} \to \mathbb{R}$ defined as follows,
	\begin{align}
		g(\mu_{X}^1, \ldots, \mu_{X}^K, \mu_{Y}^1, \ldots, \mu_{Y}^K) &=  \frac{ \sum_{k=1}^K w_k \mu_X^k \mu_Y^k -   \mu_{X,w} \mu_{Y,w}  }{\sum_{k=1}^K w_k(\mu_X^k)^2 -  \left( \sum_{k=1}^K w_k \mu_X^k \right)^2}, \nonumber \\
		& = \frac{\mbox{Cov}_w(X,Y)}{\mbox{Var}_w(X)},
	\end{align}
	with the notations $\mu_{X,w}  =  \sum_{k=1}^K w_k \mu_X^k$, $\mu_{Y,w} =   \sum_{j=1}^K w_j \mu_Y^j$,	
	$\mbox{Cov}_w(X,Y) =  \sum_{k=1}^K w_k \mu_X^k \mu_Y^k -   \mu_{X,w} \mu_{Y,w}$ and $\mbox{Var}_w(X) =  \sum_{k=1}^K w_k (\mu_X^k)^2 -   (\mu_{X,w})^2$.  The gradient $\nabla g$  of $g$, evaluated at the point $(\mu_{X}^1, \ldots, \mu_{X}^K, \mu_{Y}^1, \ldots, \mu_{Y}^K)$, is equal to 
	\begin{align*}
		\nabla g &= 
		\begin{pmatrix}
			\frac{w_1(\mu_Y^1- \mu_{Y,w})}{\mbox{Var}_w(X)} - \frac{ 2 w_1 \left( \mu_X^1 - \mu_{X,w} \right) \mbox{Cov}_w(X,Y)}{(\mbox{Var}_w(X))^2   }
			\\
			\vdots
			\\
			\frac{w_K (\mu_Y^K- \mu_{Y,w})}{\mbox{Var}_w(X)} - \frac{ 2 w_K \left( \mu_X^K - \mu_{X,w} \right) \mbox{Cov}_w(X,Y)}{(\mbox{Var}_w(X))^2}
			\\
			\frac{w_1 \left(\mu_X^1- \mu_{X,w}  \right)}{\mbox{Var}_w(X)} \\
			\vdots
			\\
			\frac{w_K \left(\mu_X^K- \mu_{X,w}  \right)}{\mbox{Var}_w(X) }
		\end{pmatrix}.
	\end{align*}
	As in the proof of Proposition~\ref{prop:mm_asympt_norm}, we get that $\sqrt{n} \left(\widehat{\beta}_1 - \beta_1 \right)  \rightsquigarrow \mathcal{N} (0,\sigma^2_{\beta_1})$ with 
	$\sigma^2_{\beta_1} = \nabla g^T \Gamma_{\mu} \nabla g$, so that 
	\begin{align}
		\sigma^2_{\beta_1} &= \frac{1}{(\mbox{Var}_w(X))^2} \sum_{k=1}^K w_k^2 \left[ \sigma^2_{X,k}  \left( \mu_Y^k- \mu_{Y,w} - 2 \beta_1 \left(\mu_X^k - \mu_{X,w} \right)\right)^2  +  \sigma^2_{Y,k}  \left( \mu_X^k- \mu_{X,w}\right)^2 \right] \nonumber  \\
		& = \frac{1}{ (\mbox{Var}_w(X))^2} \sum_{k=1}^K w_k^2 \left[ \sigma^2_{X,k}  \left(  -\beta_1 \left(\mu_X^k - \mu_{X,w} \right)  \right)^2 + \sigma^2_{Y,k}  \left( \mu_X^k- \mu_{X,w}\right)^2  \right] \nonumber \\
		&  = \frac{1}{ (\mbox{Var}_w(X))^2} \sum_{k=1}^K w_k^2    \left(\mu_X^k - \mu_{X,w}  \right)^2 \left(  \beta_1^2 \sigma^2_{X,k}  +  \sigma^2_{Y,k} \right)  
		\label{def:sigbeta1}
	\end{align}
	remarking that $\beta_1 = \mbox{Cov}_w(X,Y)/\mbox{Var}_w(X,Y)$, $\beta_0 = \mu_{Y,w} - \beta_1 \mu_{X,w}$ as well as  $\beta_0 = \mu_{Y}^k - \beta_1 \mu_{X}^k$.

\end{proof}

\begin{proof}{\em of Proposition \ref{prop:bootM}}
	
	The fact that the bootstrap estimator $\boldsymbol{\beta}^{M,*}$ is strongly consistent for $\boldsymbol{\beta}$ is a direct consequence of Theorem 3.1 in \cite{Shao_Tu_1995}, noting that 
	$$
	\widehat{\boldsymbol{\beta}}^M = g( \widehat{\boldsymbol{\mu}}_{X}^1, \ldots,  \widehat{\boldsymbol{\mu}}_{X}^K, \widehat{\mu}_{Y}^1, \ldots, \widehat{\mu}_{Y}^K)
	$$
	is a continuously differentiable function of means at $(\boldsymbol{\mu}_{X}^1, \ldots,  \boldsymbol{\mu}_{X}^K, \mu_{Y}^1, \ldots, \mu_{Y}^K)$.
	The fact that confidence sets based on the percentile approach are consistent is proved by checking the assumptions in Theorem 4.1 (iii) \cite{Shao_Tu_1995}, namely the  bootstrap estimator $\boldsymbol{\beta}^{M,*}$ is consistent, $\widehat{\boldsymbol{\beta}}^M$ is consistent (Lemma \ref{lem:mm_consist}), with asymptotic Gaussian distribution~(Proposition~\ref{prop:mm_asympt_norm}).
\end{proof}

\begin{proof}{\em of Proposition \ref{prop:bootW}}
	
	We denote by $\boldsymbol{\theta}_0 = (\boldsymbol{\beta},  \sigma_\epsilon^{2, W})$ the vector of true parameters, by $\widehat{\boldsymbol{\theta}} = (\boldsymbol{\beta}^{W},  \sigma_\epsilon^{2})$ the sequence of minimum Wasserstein distance estimators and by $\boldsymbol{\theta}^{*} = (\boldsymbol{\beta}^{W,*},  \sigma_\epsilon^{2,W,*})$ bootstrap estimators of $\boldsymbol{\theta}_0$. The vector of parameters $\boldsymbol{\theta}^{*}$ is  the minimizer of functional $\varphi_n^{*}$ defined as follows,
	\begin{align}
		\varphi_n^{*}(\boldsymbol{\gamma}, \sigma^2) &= \sum_{k=1}^K \pi_k \left[ 
		(\mu_{Y}^{k,*} - \gamma_0 - \boldsymbol{\gamma}_{-0}^\top \boldsymbol{\mu}_{X}^{k,*})^2 +  \left(\sigma_{Y,k}^* - \sqrt{\boldsymbol{\gamma}_{-0}^\top \boldsymbol{\Gamma}_X^{k,*}  \boldsymbol{\gamma}_{-0}+ \sigma^2} \right)^2
		\right].
	\end{align}
	
	We first show with arguments similar to those employed in the proof of Lemma~\ref{lem:ot_consist}, that $\boldsymbol{\theta}^{*}$ is a consistent estimator for $\boldsymbol{\theta}_0$, based on the fact that  $\varphi_n^{*}$ is a smooth function converging to $\varphi$ and the sample mean theorem for bootstrap (see for example Theorem 23.4 in  \cite{MR1652247}). Indeed, we first recall that for all $(\boldsymbol{\gamma}, \sigma^2_{\gamma}) \in \Theta$, $\varphi_n(\boldsymbol{\gamma}, \sigma^2_{\gamma}) \to \varphi(\boldsymbol{\gamma}, \sigma^2_{\gamma})$ in probability, when $n_{\min}$ tends to infinity and
	\begin{align}
		\left| \varphi_n^*(\boldsymbol{\gamma}, \sigma^2_{\gamma}) - \varphi(\boldsymbol{\gamma}, \sigma^2_{\gamma})  \right| & \leq \left| \varphi_n^*(\boldsymbol{\gamma}, \sigma^2_{\gamma}) - \varphi_n(\boldsymbol{\gamma}, \sigma^2_{\gamma})  \right| + \left| \varphi_n(\boldsymbol{\gamma}, \sigma^2_{\gamma}) - \varphi(\boldsymbol{\gamma}, \sigma^2_{\gamma})  \right|. 
		\label{ineq:phibn0}
	\end{align}
	Since the bootstrap means converge to the empirical ones, we deduce with the continuous mapping theorem that $\varphi_n^*(\boldsymbol{\gamma}, \sigma^2_{\gamma}) \to \varphi_n(\boldsymbol{\gamma}, \sigma^2_{\gamma})$ in probability, when $n_{\min}$ tends to infinity, so that $\varphi_n^*(\boldsymbol{\gamma}, \sigma^2_{\gamma}) \to \varphi(\boldsymbol{\gamma}, \sigma^2_{\gamma})$. We also have, as in~(\ref{ineq:deltaphin}), where empirical means are replaced by the bootstrap means,
	\begin{align}
		\left|  \varphi_n^*(\boldsymbol{\gamma}, \sigma^2_{\gamma}) - \varphi_n^*(\boldsymbol{\alpha}, \sigma_\alpha^2) \right| & \leq  \left(  \| \boldsymbol{\alpha} - \boldsymbol{\gamma} \|  + | \sigma_\alpha^2 - \sigma_\gamma^2 | \right)
		\sum_{k=1}^K \pi_k \left( B_n^{k,*} + A_n^{k,*} \right),
	\end{align} 
	for any $(\boldsymbol{\alpha}, \sigma_\alpha^2) \in \Theta$, with $\sum_{k=1}^K \pi_k \left( B_n^{k,*} + A_n^{k,*} \right) = O_p(1)$. 
	As a result, we deduce from Lemma~\ref{lem:ot_consist},   inequality (\ref{ineq:phibn0}) and Lemma 2.9 in  \cite{MR1315971} that
	\[
	\sup_{ (\boldsymbol{\gamma}, \sigma^2_\gamma) \in \Theta}  \left| \varphi_n^*(\boldsymbol{\gamma}, \sigma^2_{\gamma}) - \varphi(\boldsymbol{\gamma}, \sigma^2_{\gamma})\right| \to 0  \quad \mbox{ in probability.}
	\] 
	We conclude that  $\boldsymbol{\theta}^{*} \to \boldsymbol{\theta}_0$ in probability in view of Theorem 2.1 in \cite{MR1315971}.

	We now prove that $\sqrt{n} \left( \boldsymbol{\theta}^{*} - \widehat{\boldsymbol{\theta}} \right)$ and $\sqrt{n} \left( \widehat{\boldsymbol{\theta}} - \boldsymbol{\theta}_0  \right)$ have the same asymptotic distribution. By definition of $\widehat{\boldsymbol{\theta}}$ and Taylor expansion we have
	\begin{equation}
		\nabla \varphi_n(\widehat{\boldsymbol{\theta}})
		= \nabla \varphi_n(\boldsymbol{\theta}_0) +  \nabla_{00} \varphi_n(\overline{\boldsymbol{\theta}}) \left( \widehat{\boldsymbol{\theta}} - \boldsymbol{\theta}_0 \right) =  0,
		\label{pr:bw1}
	\end{equation}
	where $\overline{\boldsymbol{\theta}}$ belongs, componentwise, to the segment between $\boldsymbol{\theta}_0$ and $\widehat{\boldsymbol{\theta}}$. We have a similar expansion for boostrap estimators, 
	as well as 
	\begin{equation}
		\nabla \varphi_n^{*}(\boldsymbol{\theta}^*)
		= \nabla \varphi_n^*(\boldsymbol{\theta}_0) +  \nabla_{00}^* \varphi_n(\overline{\boldsymbol{\theta}}^*) \left(\boldsymbol{\theta}^* - \boldsymbol{\theta}_0 \right) =  0,
		\label{pr:bw2}
	\end{equation}
	where $\overline{\boldsymbol{\theta}}^*$ belongs, componentwise, to the segment between $\boldsymbol{\theta}_0$ and $\boldsymbol{\theta}^*$. 
	Combining  (\ref{pr:bw1}) and (\ref{pr:bw2}), we deduce
	\begin{align}
		\hspace{-10pt}
		\boldsymbol{\theta}^* - \widehat{\boldsymbol{\theta}}  & = \left( \nabla_{00}^* \varphi_n(\overline{\boldsymbol{\theta}}^*)\right)^{-1}  \nabla \varphi_n^*(\boldsymbol{\theta}_0) - \left( \nabla_{00} \varphi_n(\overline{\boldsymbol{\theta}})\right)^{-1}  \nabla \varphi_n(\boldsymbol{\theta}_0) \nonumber \\
		&=   \left( \left( \nabla_{00}^* \varphi_n(\overline{\boldsymbol{\theta}}^*)\right)^{-1}  - \left( \nabla_{00} \varphi_n(\overline{\boldsymbol{\theta}})\right)^{-1}  \right)  \nabla \varphi_n^*(\boldsymbol{\theta}_0) +  \left( \nabla_{00} \varphi_n(\overline{\boldsymbol{\theta}})\right)^{-1} \left(  \nabla \varphi_n^*(\boldsymbol{\theta}_0) -  \nabla \varphi_n(\boldsymbol{\theta}_0) \right).
		\label{dif:thetab}
	\end{align} 
	Noticing that $\nabla_{00}^* \varphi_n(\overline{\boldsymbol{\theta}}^*)$ and  $\nabla_{00} \varphi_n(\overline{\boldsymbol{\theta}})$ both tend in probability to the same limit  $\mathbf{H}(\boldsymbol{\beta}, \sigma^2_{\epsilon})$ and 
	we have, with similar arguments as those used in the proof of Proposition~\ref{prop:ot_asympt_norm}, that 
	$\nabla \varphi_n^*(\boldsymbol{\theta}_0)$ is $O_p(n^{-1/2})$. It can be deduced that 
	\begin{align}
		\boldsymbol{\theta}^* - \widehat{\boldsymbol{\theta}}  
		&=     \left( \nabla_{00} \varphi_n(\overline{\boldsymbol{\theta}})\right)^{-1} \left(  \nabla \varphi_n^*(\boldsymbol{\theta}_0) -  \nabla \varphi_n(\boldsymbol{\theta}_0) \right) +o_P(n^{-1/2}).
		\label{dif:thetab2}
	\end{align} 
	
	Using arguments similar to those employed in the expansion of $\nabla \varphi_n$ in the proof of Proposition~\ref{prop:ot_asympt_norm}, we make appear the difference between the bootstrap means and the empirical means or a differentiable functional of these quantities:
	
	\begin{align}
		\hspace{-10pt}
		\nabla \varphi_n^*(\boldsymbol{\theta}_0) -  \nabla \varphi_n(\boldsymbol{\theta}_0)  =
		\begin{footnotesize}
			\begin{pmatrix} 
				2 \sum_{k=1}^K \pi_k \left( (\widehat{\mu}_{Y}^k -{\mu}_{Y}^{k,*})  - \beta_0 - \boldsymbol{\beta}_{-0}^\top \left(\widehat{\boldsymbol{\mu}}_{X} -\boldsymbol{\mu}_{X}^{k,*}\right) \right) \\
				\begin{pmatrix}
					2 \sum_{k=1}^K \pi_k \left[ \left( \widehat{\mu}_{Y}^k - \mu_{Y}^k  - \beta_0 - \boldsymbol{\beta}_{-0}^\top \widehat{\boldsymbol{\mu}}_{X}^k\right)\widehat{\boldsymbol{\mu}}_{X}^k + \left(\frac{ \widehat{\sigma}_{Y,k}}{\sqrt{ \boldsymbol{\beta}_{-0}^\top \widehat{\boldsymbol{\Gamma}}_X^k\boldsymbol{\beta}_{-0} + \sigma_\epsilon^2}}  - 1 \right)  \widehat{\boldsymbol{\Gamma}}_X^k  \boldsymbol{\beta}_{-0} 
					\right]					\\ 
					-2 \sum_{k=1}^K \pi_k \left[ \left( \mu_{Y}^{k,*} - \mu_{Y}^k  - \beta_0 - \boldsymbol{\beta}_{-0}^\top \boldsymbol{\mu}_{X}^{k,*}\right)\boldsymbol{\mu}_{X}^{k,*} + \left(\frac{ \sigma_{Y,k}^*}{\sqrt{ \boldsymbol{\beta}_{-0}^\top \boldsymbol{\Gamma}_X^{k,*}\boldsymbol{\beta}_{-0} + \sigma_\epsilon^2}}  - 1 \right)  \boldsymbol{\Gamma}_X^{k,*}  \boldsymbol{\beta}_{-0} \right] \end{pmatrix} \\
				\sum_{k=1}^K  \left( \frac{\widehat{\sigma}_{Y,k}}{\sqrt{ \boldsymbol{\beta}_{-0}^\top \widehat{\boldsymbol{\Gamma}}_X^k  \boldsymbol{\beta}_{-0} + \sigma_\epsilon^2}} -  \frac{\sigma_{Y,k}^*}{\sqrt{ \boldsymbol{\beta}_{-0}^\top \boldsymbol{\Gamma}_X^{k,*}  \boldsymbol{\beta}_{-0} + \sigma_\epsilon^2}} \right)
			\end{pmatrix}
		\end{footnotesize},
		\label{def:bootgradvarphi_n}
	\end{align}
	which satisfies the central limit theorem for the bootstrap means, or the Delta method for the bootstrap estimators (see Theorem \ref{the:bmclt} in Section \ref{sec:theorems}, as well as Theorem 23.4 and Theorem 23.5 in  \cite{MR1652247}).  Consequently, $\nabla \varphi_n^*(\boldsymbol{\theta}_0) -  \nabla \varphi_n(\boldsymbol{\theta}_0)$ and $\nabla \varphi_n(\boldsymbol{\theta}_0) -  \nabla \varphi(\boldsymbol{\theta}_0)$ have the same asymptotic distribution. 
	By Slustky's theorem, the asymptotic distribution of $\sqrt{n} \left( \boldsymbol{\theta}^* - \widehat{\boldsymbol{\theta}}\right)$ is the same as the asymptotic distribution of $\mathbf{H}(\boldsymbol{\beta}, \sigma^2_{\epsilon}) \sqrt{n}  \nabla \varphi_n(\boldsymbol{\theta}_0)$, and we can conclude that $\sqrt{n} \left( \boldsymbol{\theta}^* - \widehat{\boldsymbol{\theta}}  \right)$ and  $\sqrt{n} \left( \widehat{\boldsymbol{\theta}} - \boldsymbol{\theta}_0  \right)$ also have the same asymptotic Gaussian distributions.  
\end{proof}	

\newpage
\bibliographystyle{apalike}
\bibliography{biblio_reg}

%\newgeometry{left=3cm,right=3cm, top=2cm,bottom=2cm}
\begin{sidewaysfigure}
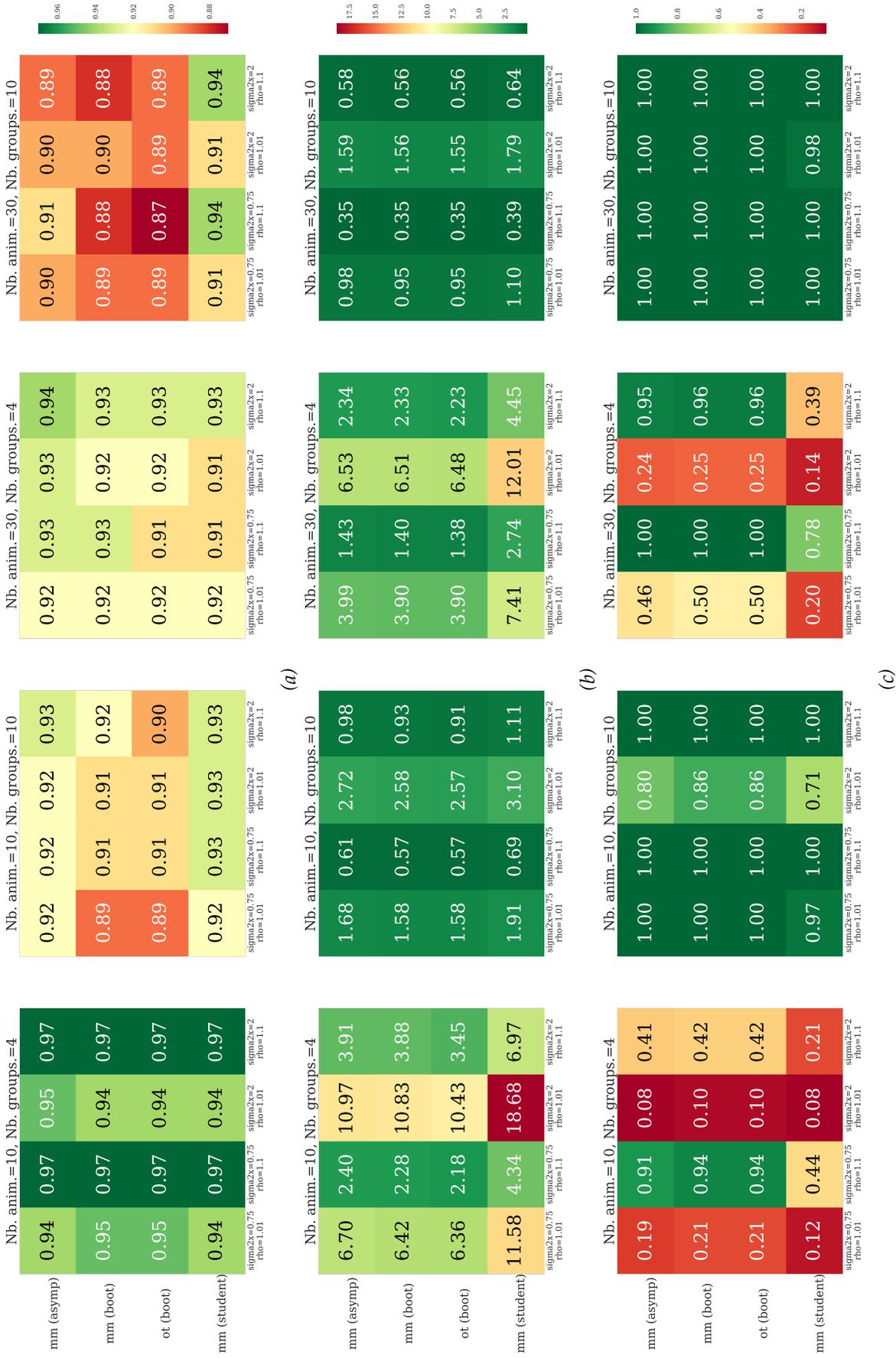

	\section{Supplementary figures}
	\centering
	%	\vspace{450pt}
	\begin{subfigure}{0.95\textheight}
		\includegraphics[clip, trim=6cm 0.8cm 5.5cm 1.1cm,width=.95\textheight]{Images/sim_cr.pdf}
		\caption{\label{fig:ci_cov_rate}}
	\end{subfigure}
	\begin{subfigure}{.95\textheight}
		\includegraphics[clip, trim=6cm 0.8cm 5.5cm 1.1cm,width=.95\textheight]{Images/sim_ci_ampl.pdf}
		\caption{\label{fig:ci_amplitude}}
	\end{subfigure}
	\begin{subfigure}{.95\textheight}
		\includegraphics[clip, trim=6cm 0.8cm 5.5cm 1.1cm,width=.95\textheight]{Images/sim_ci_power.pdf}
		\caption{\label{fig:ci_power}}
	\end{subfigure}
	\caption{\label{fig:vivo_sim_results}a) Coverage rates, b) average amplitudes, and c) powers of the confidence intervals for the estimators of $\beta_1$ obtained from $N_{sim}$ simulations for each parameter combination. The method labels on the left: "mm (asymp)" stands for the method of moments with asymptotic confidence intervals, "mm (boot)" for the method of moments with bootstrap, "ot (boot)" for the optimal transport method with bootstrap, and "mm (student)" for the naive linear regression on means approach based on Student's distribution, and "simultaneous" for the classical linear regression estimation in the case where the predictor and the predicted variable are observed simultaneously. The columns of the tables indicate simulation scenarios with different combinations of parameters:  scenario S1 with lower group overlap ($\sigma^2_{X}=0.75$) and higher signal-to-noise ratio ($\rho=1.1$), S2 with higher group overlap ($\sigma^2_{X}=2$) and higher signal-to-noise ratio ($\rho=1.1$), S3 with lower group overlap ($\sigma^2_{X}=0.75$) and lower signal-to-noise ratio ($\rho=1.01$), and S4 with higher group overlap ($\sigma^2_{X}=2$) and lower signal-to-noise ratio ($\rho=1.01$). }
\end{sidewaysfigure}

\end{document}